\documentclass[10pt]{article}

\usepackage{amsmath}
\usepackage{amssymb}

\usepackage{graphicx}

\usepackage{cite}

\usepackage{color} 
\usepackage{slashbox}

\usepackage[labelfont=bf,labelsep=period,justification=raggedright]{caption}

\makeatletter
\renewcommand{\@biblabel}[1]{\quad#1.}
\makeatother

\date{}

\begin{document}

\begin{flushleft}
{\Large
\textbf{Of lice and math: using models to understand and control populations of head lice}
}
\\
Mar{\'\i}a Fabiana Laguna$^{1}$, 
Sebasti{\'a}n Risau-Gusman$^{2,\ast}$, 
\\
\bf{1} Consejo Nacional de Investigaciones Cient{\'\i}ficas
y T{\'e}cnicas, Argentina, Centro At{\'o}mico
Bariloche, 8400 S. C. de Bariloche, Argentina
\\
\bf{2} Consejo Nacional de Investigaciones Cient{\'\i}ficas
y T{\'e}cnicas, Argentina, Centro At{\'o}mico
Bariloche, 8400 S. C. de Bariloche, Argentina
\\
$\ast$ E-mail: Corresponding srisau@cab.cnea.gov.ar
\end{flushleft}

\section*{Abstract}
In this paper we use detailed data about the biology of the head louse ({\it pediculus humanus capitis}) to build a model of the evolution of head lice colonies. Using theory and computer simulations, we show that the model can be used to assess the impact of the various strategies usually applied to eradicate head lice, both conscious (treatments) and unconscious (grooming). In the case of treatments, we study the difference in performance that arises when they are applied in systematic and non-systematic ways. Using some reasonable simplifying assumptions (as random mixing of human groups and the same mobility for all life stages of head lice other than eggs) we model the contagion of pediculosis using only one additional parameter. It is shown that this parameter can be tuned to obtain collective infestations whose variables are compatible with what is given in the literature on real infestations. We analyze two scenarios: one where group members begin treatment when a similar number of lice are present in each head, and another where there is one individual who starts treatment with a much larger threshold (`superspreader'). For both cases we assess the impact of several collective strategies of treatment.
\section*{Author Summary}
Head lice are ectoparasites that can only live on human heads, and their presence in human groups is documented since antiquity. Infestations of head lice affect primarily children, at all levels of society and most ethnic groups. Its prevalence is so high that it is an important concern for public health offices worldwide. Thus, given the long history of pediculosis and the widespread application of mathematical epidemiology to the study of various diseases, it is surprising to find that the study of pediculosis has long been neglected. In fact, most predictions about it are usually based on little more than educated guesses. We developed a mathematical model of populations of head lice based in the few detailed data available. It allows us to propose some answers for  biologically relevant questions as, for example, what is the mechanism responsible for the low levels of infestations in spite of the large number of eggs laid by female lice. We also analyze the performance of some control measures that can be adopted to eradicate head lice. Finally, we include in our model the interaction between populations to study the transmission of lice from head to head and the effect of collective control measures.

\section*{Introduction}

Of the thousands of species of blood sucking ectoparasites known as lice, only three of them infest human populations: {\it Phtirus pubis} (pubic lice), {\it Pediculus humanus humanus} (body lice), and {\it Pediculus humanus capitis} (head lice). Pubic lice are not considered a serious threat for public health because they are not known to be vectors of any diseases, and because their prevalence (defined as the proportion of infested people in a given population) is relatively low ($\approx2\%$)~\cite{chaney}. As they are mainly transmitted through sexual contact, they are often used as predictors for the presence of sexually transmitted diseases. Body lice, on the other hand, are vectors of several serious diseases such as trench fever and typhus. As infestations with body lice are most frequent in conditions of heavy crowding and poor hygiene, they have been responsible for outbreaks of typhus in times of war and in refugees camps~\cite{meinking}. After the 2nd World War, however, they have ceased to be a major public health concern.

Even though they are not known to transmit any diseases, for centuries now head lice have been a constant source of worries both for parents of infested children as for public health officials. After the 2nd World War the use of the insecticide DDT led to a temporary decrease in the prevalence of human louse, and consequently the scientific community seemed to lose interest in the study in the biology of this parasite \cite{burgess}. Resistance to insecticides and other factors, however, led to a new increase in head lice prevalence. As a result, there is a gap of almost 50 years in the few studies of the biology of the human louse. An example of this is the taxonomic status of head and body lice, which still remains unclear, in spite of several very recent studies (see, e.g.,~\cite{light,li}). 

In the 21st century the prevalence of head lice is still very high worldwide \cite{falagas}: it is not uncommon to find more than 20 \% of children infected in some schools. As a consequence, a large amount of resources is dedicated each year by governments around the world to develop new products and to design strategies for the control and prevention of the spread of head lice. To assess the impact of these measures and to be able of making meaningful predictions, one needs either detailed experiments on human subjects or a detailed knowledge of the fundamental biology of the louse. The obvious practical and moral issues of experimenting in humans have turned research to {\it in vitro} experiments, but the extrapolation of the corresponding results is far from straightforward. On the other hand, for theoretical predictions, the problem of scarcity of detailed data about the fundamental biology of the louse is compounded by the fact that in general only a small subset of these data is used, and only for qualitative reasonings.

Head lice are ectoparasites that can only live on human heads. They go through three life stages: egg, nymph and adult. The eggs, usually called {\it nits}, are glued to the hair shafts, making its removal very difficult. After leaving the egg, the nymph moults three times before turning adult. Both adults and nymphs feed on the blood of the host between 3 and 10 times a day \cite{speare06}, and can only survive a few hours without a blood meal \cite{takano03}. Transmission of lice occurs mainly from head to head \cite{canyon02}, and it has been argued that fomites may also play some role \cite{burkhart07}, but there is some controversy about this \cite{canyon10}.

Infestations of head lice affect primarily children between 3 and 11 year old, at all levels of society and most ethnic groups \cite{meinking}. Symptoms of prolonged infestations can include intense pruritus and sleeplessness \cite{meinking} and even lead to social stigmatization \cite{mclaury}. As mentioned before, the prevalence is so high that it continues to be an important concern for public health offices worldwide. 
In fact, the economical burden of treating head lice infestations has been estimated at $1$ billion \cite{hansen}.
This shows that many treatment strategies and pharmacological therapies have been proposed to control and eliminate populations of head lice. The assessment of such measures, however, is only very loosely based on biological data about the louse. And, even when the few hard data available is used, the studies analyze only simple worst- and average-case scenarios \cite{lebwohl}.

Mathematical models provide a framework into which available data can be integrated to obtain meaningful predictions about the system which is being studied. In particular, models of animal populations have a long history~\cite{watt,begon}. There are even some mathematical models of populations of some ectoparasites such as ticks~\cite{beugnet98}, fleas~\cite{beugnet04} and sea lice~\cite{revie}. Here we propose a model of populations of head lice which is based in the few detailed data available. Using both theory and computer simulations, we are able to suggest some answers for biologically relevant questions as, for example, what is the mechanism responsible for the low levels of infestations in human heads (typically $\approx10$ adult lice~\cite{meinking}) in spite of the large number of eggs laid by female lice through their adult lives. We also analyze the performance of some of the many possible control measures that can be adopted to control, or even eradicate, head lice. The model is extended to include also the possibility of interaction between populations to model the transmission of lice from head to head. In this framework, we are also able to analyze the effect of the possible collective control measures.

\section*{Materials and Methods}

To build our model of head lice we use a mixed Leslie-Lefkovitch matrix approach. The Leslie matrix approach~\cite{leslie,caswell} implies characterizing the population by the age of the individuals, whereas with the Lefkovitch matrix~\cite{lefkovitch} age is disregarded in favour of life stages. In our case, we use both life stages and the age of the individual inside every life stage. We consider five life stages: egg, nymph before the first, second and third moults, and adult. The maximal durations of each stage are, respectively, $n_e$, $n_1$, $n_2$, $n_3$, and $n_a$, measured in days. Therefore, the vector that characterizes the population has $n_e+n_1+n_2+n_3+n_a$ components. The first component corresponds to the total number of eggs laid in one day, the second corresponds to the total number of one day old eggs, etc. Component $n_e+1$ corresponds to the number of nymphs that just hatched, component $n_e+2$ to one day old nymphs, and so on. In principle, transitions can occur from any day in any stage, to the first day in the following stage. 

We have built two matrices, corresponding to two different sets of data. Unfortunately, the detailed data necessary to build the matrix is available only for body lice \cite{evans}. In that work, Evans and Smith (hereafter ES) have measured the length of the five stages, as well as the fecundity for a large number of individuals, and these measurements are one of our sets of data. Recently, Takano-Lee and co workers (hereafter TL) have succeeded in rearing head lice both in vivo and in vitro and have obtained reasonable accurate values for many biological parameters~\cite{takano03}. But they only provide mean values and dispersion for the mortality in the adult stage, which does not allow for a reconstruction of the full survival curve. To obtain the required detailed data, we note first that the adult survivorship data of Evans and Smith can be very well fitted by a Weibull distribution \cite{pinder} (as happens with other insects). Therefore we have also proposed a Weibull distribution for head lice that is compatible with the data of \cite{takano03}. Table 1 shows the values we have used for the most important vital parameters (for the rest of them, see Appendix A).

\begin{table}[ht!]
\centering
\caption{Summary of the average vital parameters of {\it Pediculus humanus capitis}. In the first five columns the duration of each stage is given in days. The last column indicates the average number of eggs. For each model (TL and ES) the first line gives the average value and the second the minimal and maximal values. The standard error of the last figure of each average is given between parenthesis.}
\begin{tabular*}{\textwidth}{@{\extracolsep{\fill}}|c|c|c|c|c|c|c|}
\hline
 &hatching& 1st moult& 2nd moult &3rd moult&adulthood&eggs\cr
\hline \hline
TL \cite{takano03}
&$8.4(1)$&$3.0(0)$&$5.2(1)$&$8.0(0)$&$20.2(1.4)$&$4.9(2)$\cr \hline
&$7-11$&$3-4$&$5-7$&$8-9$&$12-31$&$1-6$\cr \hline
ES \cite{evans}
&$8.02(1)$&$5.23(4)$&$8.63(3)$&$12.81(4)$&$17.58(46)$&$4.9(2)$\cr
\hline
&$5-12$&$4-7$&$8-10$&$12-14$&$1-46$&$3-10$\cr
\hline
\end{tabular*}
\label{table1}
\end{table}

Given an initial population vector $P_0$, the population at day $n$ is given by $P_n=M^n P_0$, where $M$ is the Leslie-Lefkovich matrix obtained using TL or ES data. The long time behaviour of the population is given by $\lambda_1$, the largest eigenvalue of $M$: if $\lambda_1>1$ there is exponential growth whereas if $\lambda_1<1$ there is extinction, regardless of the initial condition. But note that components of $P_n$ give {\it average} values for the number of insects at each stage of development. In a given population the vital parameters of each insect are stochastic parameters, whose average values are given by the matrix $M$. Thus, in a finite population there is the possibility of {\it stochastic} extinctions even if $\lambda_1>1$. Evidently stochastic extinctions become exponentially less probable as $\lambda_1$ is made larger. Another important difference for finite populations is that they can become extinct in finite time whereas in the matrix population approach this extinction time is infinite, because non-zero components $p_n$ remain greater than $0$ at all times.

To go beyond average values, which are the only quantities that can be calculated with the matrix approach, we have resorted to agent-based simulations.  They consist of populations of `agents' (i.e. head lice) which evolve stochastically through the different life stages according to their vital parameters. Each vital parameter of each agent is drawn from an exponential distribution with the same average as that given by the corresponding component of the evolution matrix. The algorithm is explained in detail in Appendix A. Both in our simulations and in the matrix population formalism we have assumed that lice do not interact. That is, the vital parameters of each agent do not depend on the total number of individuals. As blood is readily available for head lice, it is reasonable to assume that lice do not compete for the resources, if populations are not very large. 

The results we obtained with both, TL and ES data sets, have only small differences. Consequently, in the following sections the description of results applies to both group of data unless otherwise stated. Moreover, to avoid an unnecessary duplication of figures, we only show the ones corresponding to TL data. The figures drawn using ES data are given in Appendix B.

\section*{Results}

\subsection*{Single population dynamics}

For the two sets of parameters used, the {\it intrinsic growth rate}~\cite{allman}, i.e. the largest eigenvalue of the corresponding evolution matrix, is: $\lambda_1=1.12$ for the ES model and $\lambda_1=1.133$ for the TL model. Fig. \ref{fignatevol} shows the average evolution of a population initiated by a 10-day old single female (i.e. a female which has undergone her last moult 10 days ago). The time-dependence of the population is calculated using evolution matrices and by means of numerical simulations. 
In the case of the simulations, we plot the average values obtained from 1000 realizations which start with the same initial condition (one 10-day old female).
The figure shows that there is a very good agreement between theory and the average of simulations, which confirms that the algorithm is a stochastic version of the model.  
We also studied the evolution of colonies starting with different initial conditions, and found no substantial differences with the case shown in Fig.~\ref{fignatevol}.
In all our calculations we have assumed that the first female does not need to be fertilized by a male: it has been reported that females can lay eggs during several days after being fertilized~\cite{bacot}, and it has even been suggested that a single mating could be enough to achieve lifetime fertility \cite{maunder93}. The figure shows that after the first month the number of adults grows rapidly from hundreds to thousands of individuals. In real populations living in human heads, however, it is well known that the average number of live lice is typically about $10$~\cite{meinking} (although there are records of individuals with hundreds of adult lice~\cite{speare02,mumcuoglu2}).

\begin{figure} [!ht]
\centerline{\includegraphics[width=12cm,clip=true]{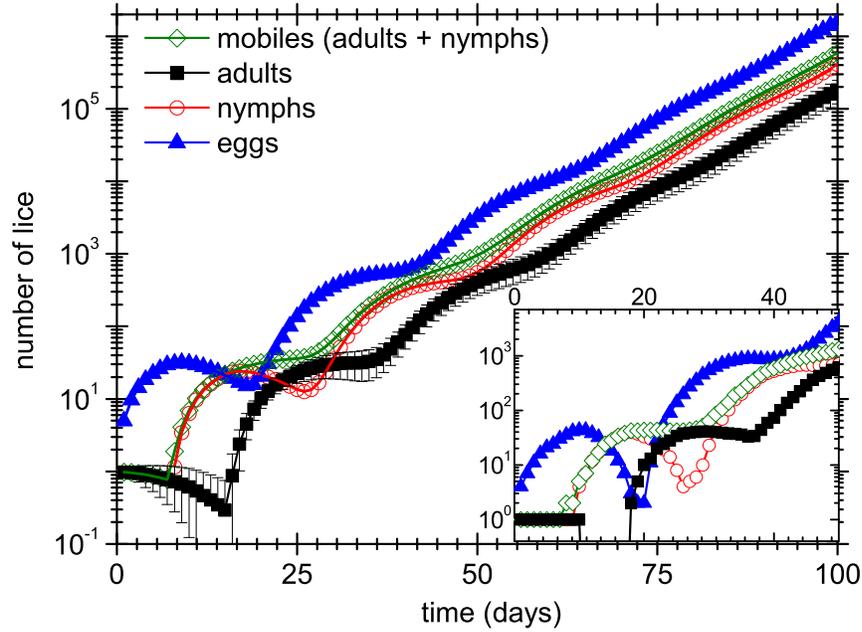}}
\caption{Average number of lice of a colony that is started at day $0$ by a female that had her last moult 10 days before. Symbols represent averages taken over $1000$ populations whereas full lines represent the theoretical predictions. The inset shows the first days of one of these populations. Here, as in the rest of the figures, the error bars represent the standard error of the mean. Data whose error bars are not shown have standard errors lower than symbol size.} 
\label{fignatevol}
\end{figure}

The strategies by which hosts control their populations of ectoparasites can be divided into three classes \cite{krasnov}: physically avoiding the parasite, exterminating it, and minimizing the harm done by it. This last line of defence is in fact an indirect means to control the population of the ectoparasite: for example, immune response could decrease the amount of blood sucked, or make it less beneficial, thereby reducing the lifespan or the reproductive success of the parasite.  We consider here two of the possible mechanisms that have been suggested as responsible for the small sizes of the lice populations in human heads. One is the triggering of some modification of the hosts blood~\cite{meinking}, which would in turn lead to a reduced fertility of female lice. 

With our model it is possible to estimate how much certain vital parameters should change to achieve an effective control of population growth. Mathematically, the critical value for a given parameter is defined as the value for which the evolution matrix has $\lambda_1=1$. In other words, the critical value separates the parameter region for which the population increases exponentially ($\lambda>1$) from the region of parameters for which it becomes extinct ($\lambda<1$). We have found that the critical value for the number of daily eggs laid by each female is approximately 1 egg per week (both for TL and ES data), which is much less than the 'natural' value (see Table 1). Furthermore, Fig. \ref{figgrowth} shows that, when considered as a function of the daily number of eggs, $\lambda_1$ grows rapidly in the vicinity of the critical value.
Thus, even a small increase on the critical number of daily eggs produces a significant intrinsic growth of the population. Studies carried out with lice fed on rabbit blood \cite{benyakir, mumcuoglu} have shown that, even when rabbits are specifically immunized (which should have a stronger effect than an spontaneously triggered immune response), the effect on female lice is to diminish the number of eggs laid to 1 or 2 per day, which is far from the critical value. This suggests that it is unlikely that the control of the lice population is done through a lowering of the reproductive success of female lice.

\begin{figure} [!ht]
\centerline{\includegraphics[width=14cm,height=10cm,clip=true]{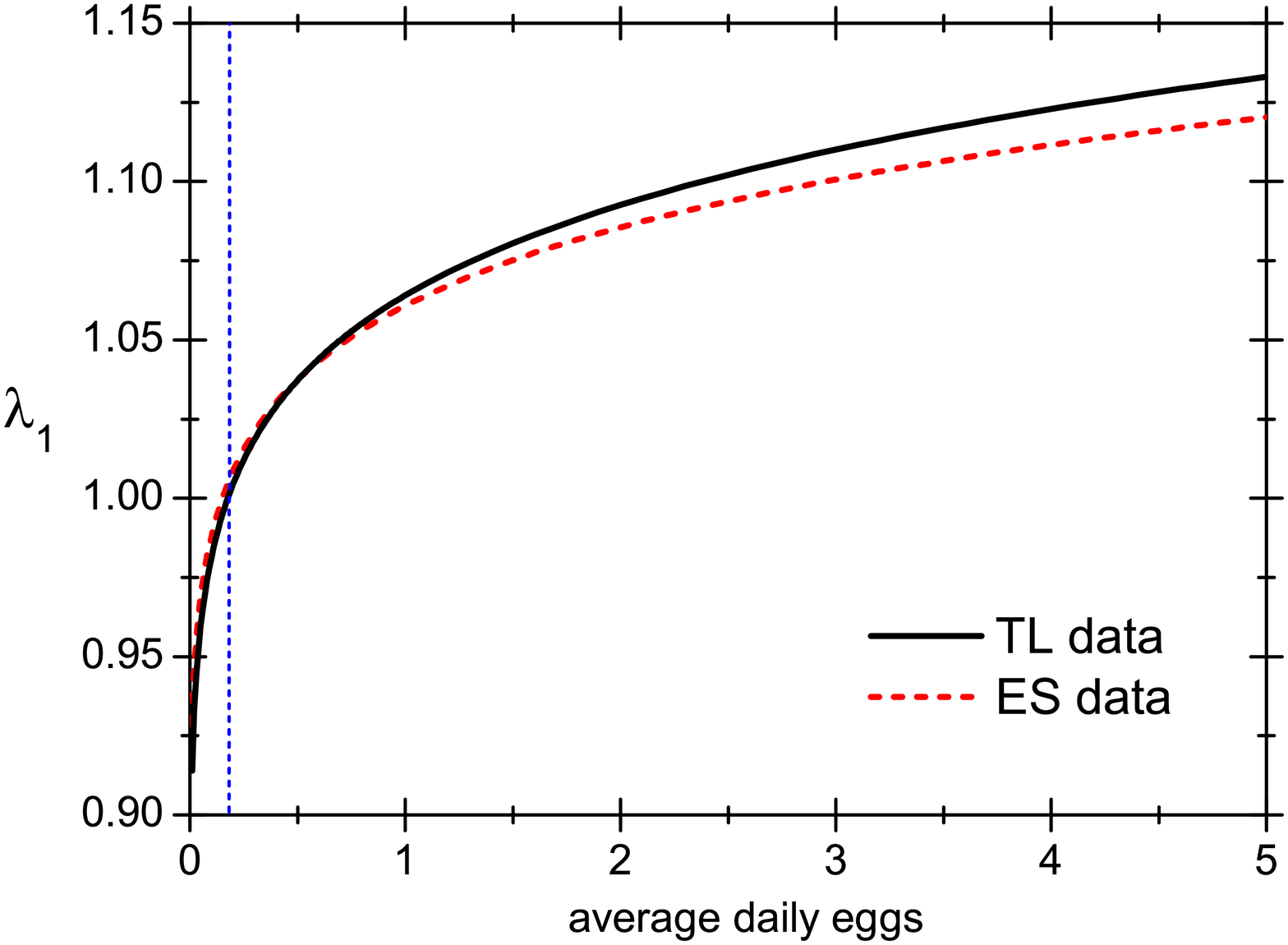}}
\caption{Intrinsic growth rate for a lice population as a function of the average of daily eggs laid by a single female. The rest of the parameters that define the population are taken from \cite{takano03} (Takano-Lee curve) and \cite{evans}(Evans/Smith curve). The vertical line shows the position of the critical value. } 
\label{figgrowth}
\end{figure}

Another possible explanation for the usually low number of lice observed is self-grooming of the host~\cite{buxton41}, which includes such activities as combing, scratching, washing, and any other action that might disturb the natural habitat of lice. Note that, in principle, these are activities that are not consciously aimed at the eradication of lice (conscious strategies for eradication lice will be treated in the next subsection). We assume that, each day, grooming eliminates a certain average fraction of all mobile lice (which are defined as the sum of nymphs and adults). It is also assumed that grooming does not remove any eggs because they are strongly attached to hair shafts by their glue \cite{burgess10}. We find that the critical percentages of mobile lice that must be removed each day to guarantee eventual extinction of the colony are $17 \%$ and $15 \%$ for TL and ES data respectively. These values seem at first sight to be rather low: extinction of the population is guaranteed if, on average, 1 out of 6 lice is removed (or killed) every day. One could suspect that, at such low levels of grooming, extinction only happens after a very long time. Figure 3 shows that this is not the case because extinction times drop sharply for grooming efficiencies larger than the critical. Furthermore, even when grooming is not effective enough to cause an extinction of the population, it can slow down its evolution, as the inset in Fig.~\ref{figgroomin} shows. As an example, note that when the efficiency of grooming is $\approx 10 \%$ it takes an additional week for the population to have 15 mobile lice, respect to the no grooming case.

These results suggest that it is more probable that an effective control of the population is achieved by killing or removing lice (i.e., by grooming) than by reducing the reproductive success of female lice (which could happens through some modification of the host blood). 

\begin{figure} [!ht]
\centerline{\includegraphics[width=15cm,height=10cm,clip=true]{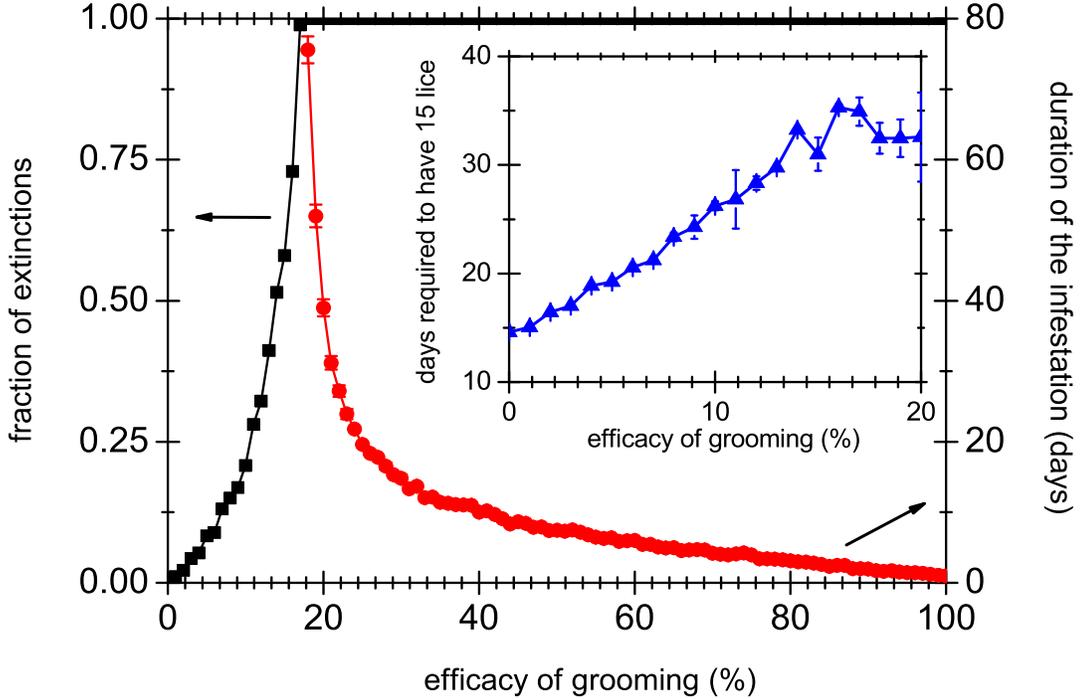}}
\caption{Fraction of extinctions (squares), and average duration of the infestation for the extinct populations (circles), as a function of the efficiency of grooming. Averages were taken over 1000 realizations, with a limit time of 500 simulation days (i.e. we only count extinctions happening within the first 500 days). The critical grooming efficacy is defined as the value at which the fraction of extinctions reach the unity. In this plot, this happens for efficacies closer to  $17 \%$. Inset: Number of days required to reach a population of $15$ lice as a function of the efficacy of grooming.} 
\label{figgroomin}
\end{figure}

\subsection*{Evaluating control measures}

In the long history of lice infestations in humans many elimination strategies and remedies have been used. This has generated a flood of tests which try to assess the efficacy of the different strategies and chemical products, which in turn has generated a large number of review articles which compare the results of different tests \cite{lapeere}. In most cases the trials consist on the application of a given treatment to a batch of infected people (mostly children) and then the fraction of `cured' people is recorded. One problem is that there seems to be no agreement on the very definition of `infestation', at least on the practical side. Infestations can be defined as the presence of live lice, or lice and eggs. This is compounded by the fact that there is no infallible method to detect head lice and/or eggs. 

For trials of chemical products, a different approach consists of applying the substance to a population of lice placed in an artificial environment, and then recording the fraction of dead insects \cite{heukelbach}. In this case, however, it is not clear how to link that number with an assessment of the recommended treatment to eliminate lice, or how to propose an effective treatment. A detailed mathematical model can provide such a link.

In the rest of this paper we use a definition of a {\it treatment} as a strategy to eradicate head lice, consisting in a series of {\it applications}. An application is defined simply as something that is done at some definite time to kill as many head lice as possible. Its efficacy is defined by the percentage of lice that are effectively eliminated. In principle, one could assume that the effect of the application is different for each life stage of the insects, or even that it depends on their age. This is represented by a diagonal matrix, $T$, of the same dimensions as the evolution matrix $M$, whose non-zero components give the fraction of insects of each age that survive the treatment. But most experimental studies of topical treatments only assess their effect on adult lice (pediculicidity) and on eggs (ovicidity)~\cite{stiche}. For this reason we have only considered matrices $T$ whose elements for the egg stage are all equal to the proposed ovicidal activity of the treatment, whereas the rest of the elements are all equal to the pediculicidity.  In other words, we assume that each treatment is defined by only two numbers, the pediculicidity and the ovicidity, as is done in most clinical trials.  We have also included a detection threshold for the start of the treatment. It is defined as the number of mobile lice necessary for a parent to notice that his/her child has an infestation, or to cause an itching feeling. In all the figures of this subsection this threshold is set to 15 mobile lice.
In Ref. \cite{meinking} it is stated that it usually takes several weeks for an individual to start itching the first time he or she has lice. Then, the threshold we choose is the number of mobile lice present in a head 3 weeks after the start of the infestation, if we consider a grooming efficacy of $5\%$ (see the inset of Fig. \ref{figgroomin}). This grooming efficacy is kept constant in the rest of the simulations shown in this paper.

To specify a treatment one needs not only the frequency of the applications but also a criterion for stopping them. In this regard we classify treatments as {\it systematic} and {\it non systematic}. For the former it is assumed that the treatment is applied regardless of the state of the lice colony. In other words, it is applied at least until the whole colony has been eliminated. 
This would correspond to strictly follow the suggested treatment, in terms of number and frequency of applications, without using a personal criterion for deciding whether the colony has already been exterminated.
Non systematic treatments are defined as those where the stopping of the treatment depends on the state of the colony. 
This models the situation of parents using their own discretion to decide when the infestation is over.
In particular, we consider that systematic treatments depend only on one parameter, $\Delta t_{a}$, which is the time elapsed between two successive applications. Moreover, non systematic treatments depend not only on $\Delta t_{a}$ but also on the parameter $\lambda_{end}$, defined as the threshold to stop the treatment: if the number of mobile lice is smaller or equal than $\lambda_{end}$ the treatment ceases to be applied. This is intended to model the fact that it is very difficult to detect every mobile lice in a given head, and thus if the number of mobile lice is sufficiently small, they will not be detected, the head would be assumed to be clean of lice, and therefore the treatment will be stopped. For all treatments, we define the {\it duration of the treatment} as the time elapsed between the first application and the extinction of the colony.

Simulations show that, for all treatments, there is a critical value of the application efficacy: below it, the probability that the treatment succeeds in exterminating the lice colony is practically 0 whereas above it it is practically 1 (see Fig.\ref{figprobdur} (A)). Interestingly, comparing both panels of Fig. \ref{figprobdur}, we observe that even though larger values of $\lambda_{end}$ give longer infestations, the critical application efficacy does not change with $\lambda_{end}$. To calculate this critical value for systematic treatments that are applied daily, the equation $\lambda_1=1$ must be solved, where $\lambda_1$ is the largest eigenvalue of the matrix $M \cdot T$. If the treatment is applied once every $n$ days the procedure is the same, but replacing $M \cdot T$ by $M^n \cdot T$. Results are presented in Table~\ref{table2}, where the pediculicidity necessary to guarantee the extinctions is given as a function of ovicidity and frequency of the treatment. Ovicidal efficacy however, is difficult to measure both in vivo and in vitro\cite{sonnberg}, and therefore the available estimates are not very reliable. Moreover it is generally agreed that the ovicidal efficacy of most products is rather low. For this reason, in Table 2 we have only included a few values of ovicidity and in the rest of the paper we have arbitrarily fixed this number at a value of $10 \%$.

\begin{table}[ht!]
\centering
\caption{Daily fraction of mobile lice that should be eliminated to cause an extinction for treatments that are applied once every $n$ days (rows). Columns indicate the ovicidity $o$, i.e., the fraction of killed eggs. The last column gives the critical fraction of eggs ($f_{eggs}$) that have to be eliminated to guarantee the extinction of the population, when no action is taken against mobile lice. All fractions are given for both TL and ES data.}
\begin{tabular*}{\textwidth}{@{\extracolsep{\fill}}|c|c|c|c|c|c|c|c|c||c|c|}
\hline
\backslashbox{$n$}{$o$}&\multicolumn{2}{c|}{$0$}&\multicolumn{2}{c|}{$0.1$}&\multicolumn{2}{c|}{$0.3$}&\multicolumn{2}{c||}{$0.5$}&\multicolumn{2}{c|}{$f_{eggs}$}\cr
\cline{2-11}
&TL&ES&TL&ES&TL&ES&TL&ES&TL&ES \\
\hline \hline
$1 d$&$0.174$&$0.146$&$0.126$&$0.11$&$0.022$&$003$&$0$&$0$&$0.34$&$0.367$\cr \hline
$2 d$&$0.318$&$0.271$&$0.277$&$0.24$&$0.18$&$0.164$&$0.054$&$0.07$&$0.57$&$0.6$\cr \hline
$3 d$&$0.439$&$0.379$&$0.404$&$0.351$&$0.319$&$0.284$&$0.204$&$0.194$&$0.728$&$0.76$\cr \hline 
$4 d$&$0.539$&$0.471$&$0.509$&$0.447$&$0.436$&$0.388$&$0.335$&$0.307$&$0.832$&$0.865$\cr \hline
$7 d$&$0.757$&$0.68$&$0.737$&$0.663$&$0.688$&$0.622$&$0.621$&$0.566$&$0.957$&$0.989$\cr \hline
$10 d$&$0.873$&$0.808$&$0.862$&$0.798$&$0.834$&$0.771$&$0.792$&$0.732$&$1$&$1$\cr \hline
\end{tabular*}
\label{table2}
\end{table}

Table~\ref{table2} shows that the number of lice that have to be eliminated at each application of the treatment is relatively low, provided that the treatment is systematic. For example, if the treatment is applied every 2 days, eliminating one out every 3 insects guarantees the {\it eventual} extinction of the population. 
But two questions arise. The first is: after how many days since the treatment is begun will the population really disappear? This is partially answered in Fig.~\ref{figprobdur}(B), where we show the average durations for different treatments. As mentioned before, in the upper panel of this figure we also plot the probability of lice colony extinction as a function of the efficacy of the pediculicidal, for an ovicidity of $10 \%$. Two systematic treatments are shown, one corresponding to daily applications ($\Delta t_{a}=1$) and the other  which is applied every 4 days ($\Delta t_{a}=4 $). We also plot the results obtained for three non systematic treatments, which differ in the value of $\lambda_{end}$. We observe that the critical efficacy - defined as the value of efficacy above which the probability of elimination of the colony is near to one - strongly depends on the parameter $\Delta t_{a}$, but it is almost independent of $\lambda_{end}$.
Moreover, note that a treatment which is applied systematically every 4 days (squared symbols) would take more than 30 days (25 for ES data) to achieve total extinction of the population if it eliminates $60 \%$ of the mobile lice each time. But if the same treatment is used with a perfect pediculicidal (i.e., $\Delta t_{a}=4 $ and $100 \%$ of efficacy) it would take only 10 days, on average, which is equivalent to 3 applications of the treatment. We are assuming here that the treatment is perfectly systematic, that is, that it is applied with a given fixed frequency and for a period of time that assures the extinction of the population. 
We are aware that this is a very unrealistic assumption, because it implies that one can detect every live lice and every egg, which is well known to be far from true~\cite{mumcuoglu01}. 
Seen from another angle, a systematic treatment assumes that parents will continue applying it even if no lice are detected, which is also unrealistic.
This leads to the second question: What changes if we relax the assumption of systematicity?

\begin{figure} [!ht]
\centerline{\includegraphics[width=10cm,clip=true]{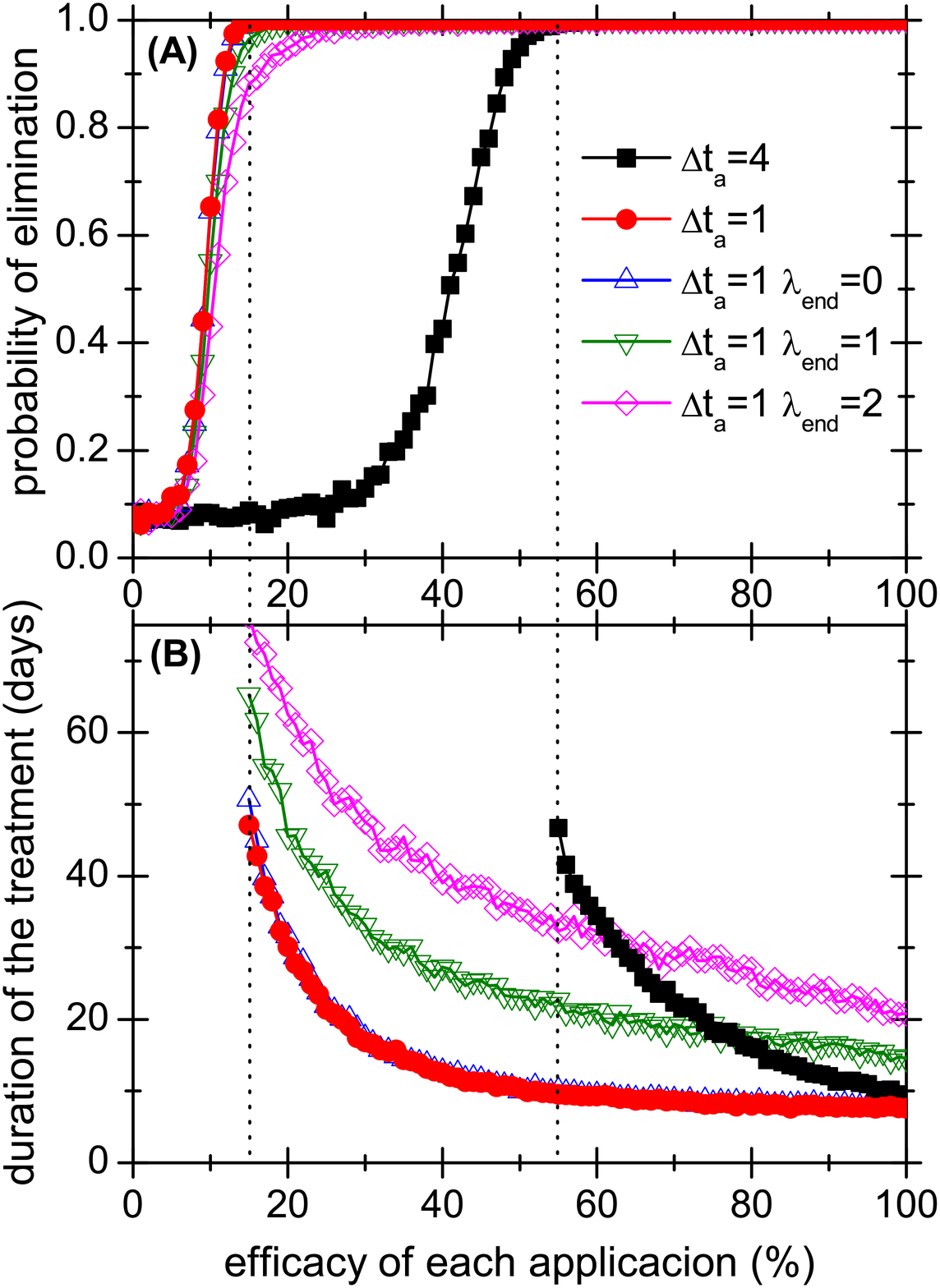}}
\caption{Results of applying different treatments to cure a head lice infestation. The upper panel shows the probability that the infestation is cured (i.e. that all head lice and eggs are eliminated) and the lower panel shows the duration of the treatment, when it is successful. Both variables are plotted as functions of the fraction of lice that are eliminated by each application of the treatment for a fixed ovicidity of $10\%$. The limit time in our simulations to allow for the extintion of the colony was $500$ days of simulation time. Dotted vertical lines indicate the critical efficacy of the treatments.} 
\label{figprobdur}
\end{figure}

\begin{figure} [!ht]
\centerline{\includegraphics[width=10cm,clip=true]{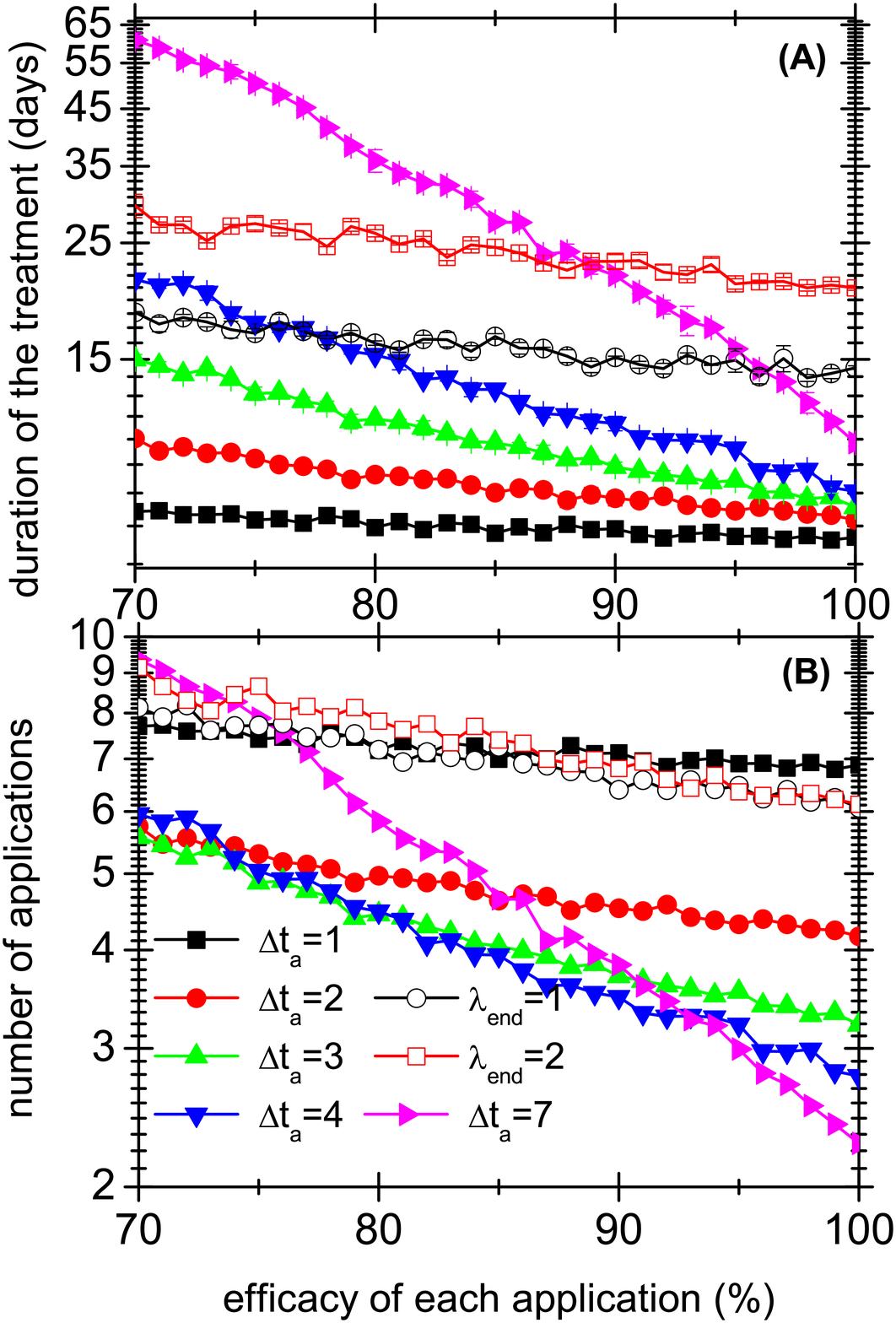}}
\caption{Comparison between the average durations (panel A) and between the number of applications (panel B) for several treatments, when they are successful, as a function of the fraction of lice eliminated by each application. Full symbols indicate systematic treatments, and $\Delta t_{a}$ gives the number of days between applications. Empty symbols correspond to a daily application ($\Delta t_a=1$) of non systematic treatments which are stopped when the mobile lice remaining in the population are less of equal than $\lambda_{end}$.} 
\label{figcompatreat1}
\end{figure}

\begin{figure} [!ht]
\centerline{\includegraphics[width=10cm,clip=true]{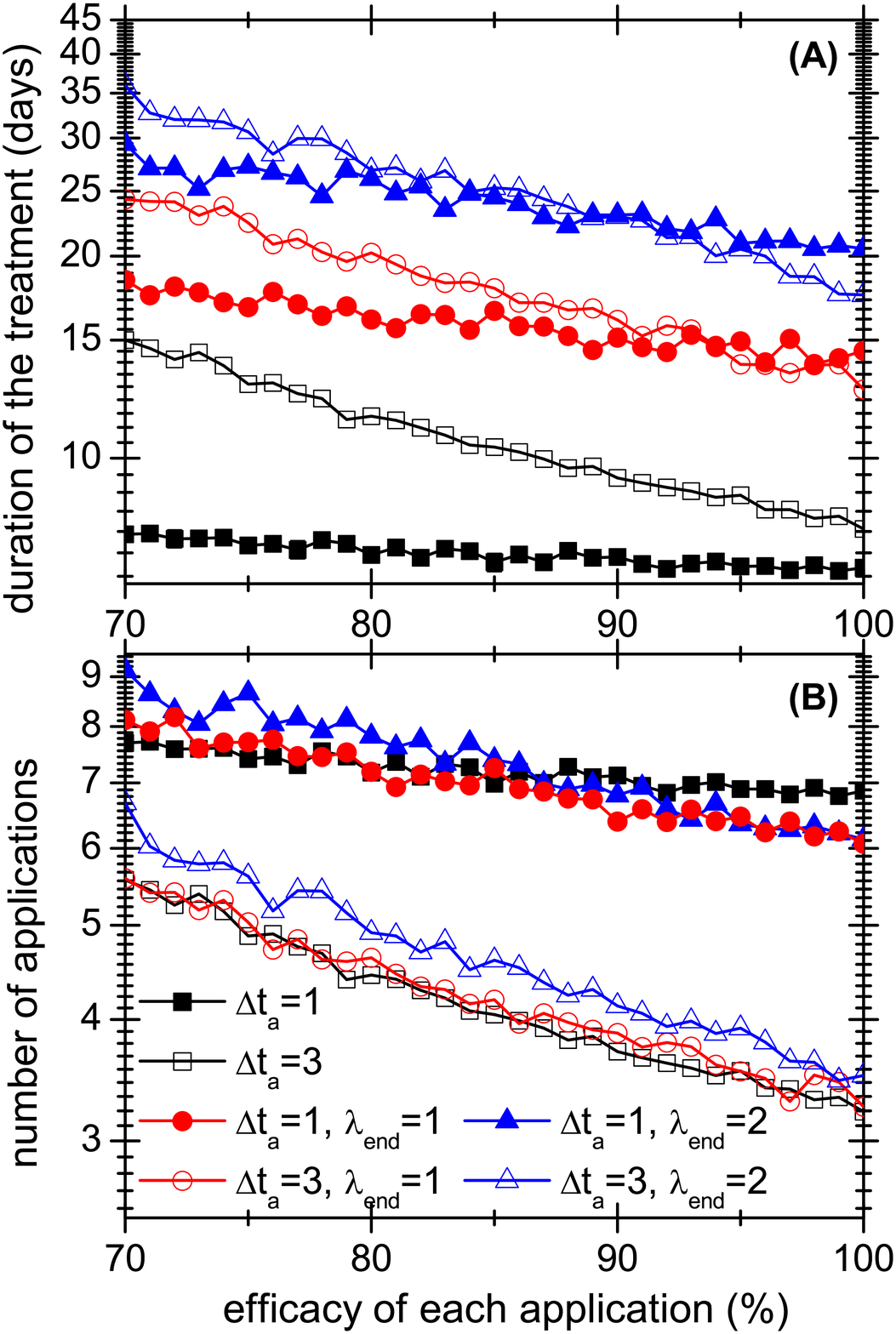}}
\caption{Comparison between the average durations (panel A) and between the number of applications (panel B) for several treatments, when they are successful, as a function of the fraction of lice eliminated by each application. Squared symbols are systematic treatments, whereas the rest are non systematic ones. Full symbols correspond to daily applications that are stopped when less than $\lambda_{end}$ mobile lice remain in the population. Empty symbols correspond to an application every 3 days.} 
\label{figcompatreat2}
\end{figure}

Turning now to non systematic treatments, we assume that the threshold $\lambda_{end}$ refers exclusively to mobile lice, because eggs are not only difficult to detect, but it is usually not easy to distinguish between live and dead eggs. The effect of using different threshold values and several application frequencies are shown in Figs.~\ref{figprobdur}, ~\ref{figcompatreat1} and ~\ref{figcompatreat2}, which present the results of $1000$ simulations of the model. In Figs.~\ref{figprobdur}(B), ~\ref{figcompatreat1}(A) and ~\ref{figcompatreat2}(A) we plot the average duration of the treatments. It is interesting to compare these times for systematic and non systematic treatments: from Fig.~\ref{figprobdur}(B) we see that if the applications have a reasonable efficacy (i.e. when the daily fraction of killed lice is larger than $\approx 80 \%$), applying the treatment every day as long as there is at least one mobile lice ($\Delta t_{a}=1, \lambda_{end}=1$) is less effective than applying it systematically every 4 days ($\Delta t_{a}=4$). 

It is well known that almost all anti-lice treatments are a nuisance because chemical products can be very aggressive and have side effects, and combing can be tiresome and stressing, both for the patient as for the comber, resulting in a loss of efficacy. Thus, a relevant quantity is the number of times that the treatment is applied in each strategy. This information is plotted in Figs.~\ref{figcompatreat1}(B) and ~\ref{figcompatreat2}(B). Fig.~\ref{figcompatreat1} shows that even though the durations of treatments with different frequencies can differ markedly, in terms of number of applications the differences are much less significant. Moreover, Fig.~\ref{figcompatreat2} shows that the curves for treatments with the same frequency but different thresholds are roughly parallel. This means that increasing the threshold $\lambda_{end}$ multiplies the duration of the treatment by a number that, at least in the interval shown, is independent of the efficacy of the applications. Interestingly, the number of applications depends only very weakly on $\lambda_{end}$.

The figures allow one to assess the efficacy of the different strategies. Not surprisingly, the time it takes for a treatment to be effective is positively correlated with the total number of applications of the strategy. But the advantage of a detailed model is that it gives less intuitive predictions. For example, Figs.~\ref{figcompatreat1} and ~\ref{figcompatreat2} show that strategies that are applied every 3 days or less often require almost the same number of applications. It also confirms the results of \cite{lebwohl}, who showed that in the worst case, three applications every 7 days were enough to eliminate all lice, if the treatment has $100 \%$ efficacy against mobile lice.

\subsection*{Interaction between populations}

To model the interaction between colonies of head lice, we assume that there is a probability per day $p_t$ that each mobile lice transfers (i.e. migrates) to another head. For simplicity, we have assumed random mixing for the human populations considered, i.e. a louse on a given head can migrate to any other head in the population. This parameter depends both on the behavioural and kinetic features that affect the movement of lice (as, for example, the probability of `catching' a passing hair shaft), as well on the social factors that can make heads come together (playing together, sharing the same bed, etc.). Even though there are some studies that try to quantify the mechanical aspects of the transmission of head lice \cite{canyon02,takano05}, many other factors remain whose quantification seems very difficult (many of them involving the various ways of interaction among school-aged children). This makes it almost impossible to estimate $p_t$ from first principles.

To have an idea of the order of magnitude of $p_t$ we have used a different method. Recently, a mathematical model of pediculosis~\cite{stone} which treated head lice infestations as `infections' (i.e. the number of lice was not taken into account) determined that the number of secondary cases caused by an `infected' individual, usually called $R_0$~\cite{anderson}, was slightly larger than $1$. In terms of colonies of head lice, this is equivalent to saying that, on average, during the whole life cycle of a colony, on average one adult female lice migrates to a different head to initiate a new colony. If we assume that the average duration of an infestation is 2-3 weeks, and that the average daily number of adult female lice is 1, the resulting daily migration probability should be between $p_t=.02$ and $p_t=.07$. As this gives only a rough estimate of the range of `reasonable' values for $p_t$, we have performed simulations using different values of $p_t$, for two different scenarios detailed below. To avoid an unbounded growth of the number of lice, we have assumed that at some point every infested individual starts some form of treatment. Different surveys have shown that this is indeed the case in most schools~\cite{speare10}. In the first scenario we have considered a population of 20 heads and we have assumed that in every infested head any louse can jump to any other head, with probability $p_t$. The treatment available to the population has an efficacy of $80 \%$ and is systematically applied every 4 days until no lice (mobiles or eggs) remain on that head. We have assumed that the detection threshold for the start of the treatment, i.e. the number of mobile lice necessary for a parent to notice that his/her child has an infestation, or to cause an itching feeling, is a number randomly distributed between 10 and 20. Scenario 2 is obtained from scenario 1 by assuming that one individual has a much larger threshold (100 mobile lice) for the start of the treatment. This choice is motivated by the fact that, even though in most infestations only $\approx 10$ lice are present, it is not uncommon to find a few children with much more acute infestations~\cite{speare02,mumcuoglu2}. As far as we know, no plausible biological explanation has been put forward to account for this, thus it seems reasonable to assume that such heavy infestations result from a delay in the beginning of the treatment.

\begin{table}[ht!]
\centering
\caption{Result of the average of several variables over 1000 runs of collective infestations for scenario 1 (group of 20 heads with thresholds for treatment beginning randomly chosen between 10 and 20 mobile lice). In the first column the probability of transfer is indicated. Second column gives the average total daily number of mobile lice involved. The third is the prevalence, i.e., the proportion of infested heads. Fourth column indicates the average daily number of lice transferred from one head to another. The fifth and sixth columns give the average and median duration of collective infestations, respectively.}
\begin{tabular*}{\textwidth}{@{\extracolsep{\fill}}|c|c|c|c|c|c|}
\hline
$p_t$&total&prevalence&total&average duration&median duration\cr
&mobile&&transferred&of infestation&of infestation\cr
\hline \hline
$0.010$&$3.77$&$4.8 \%$&$0.14$&$28.88$&$24$ \cr \hline
$0.050$&$4.91$&$6.3 \%$&$1.5$&$39.92$&$28$ \cr \hline
$0.075$&$7.82$&$9.1 \%$&$8.68$&$68.43$&$32.5$ \cr \hline
$0.100$&$13.07$&$14.1 \%$&$46.21$&$149.15$&$40$ \cr \hline
\end{tabular*}
\label{table3}
\end{table}

\begin{table}[ht!]
\centering
\caption{Result of the average of several variables over 1000 runs of collective infestations for scenario 2 (almost identical to scenario 1, but with one individual having a treatment threshold of 100 mobile lice). In the first column the probability of transfer is indicated. Second column gives the average total daily number of mobile lice involved. The third is the prevalence, i.e., the proportion of infested heads. Fourth column indicates the average daily number of lice transferred from one head to another. The fifth and sixth columns give the average and median duration of collective infestations, respectively.}
\begin{tabular*}{\textwidth}{@{\extracolsep{\fill}}|c|c|c|c|c|c|}
\hline
$p_t$&total&prevalence&total&average duration&median duration\cr
&mobile&&transferred&of infestation&of infestation\cr
\hline \hline
$0.010$&$23.59$&$7.6 \%$&$1.89$&$72.93$&$73$ \cr \hline
$0.050$&$30.48$&$19.4 \%$&$22.24$&$130.65$&$111$ \cr \hline
$0.075$&$39.35$&$28.2 \%$&$83.80$&$243.47$&$171$ \cr \hline
$0.100$&$51.89$&$38 \%$&$277.37$&$435.19$&$310.5$ \cr \hline
\end{tabular*}
\label{table4}
\end{table}

In Tables~\ref{table3} and~\ref{table4} we show the result of the average of several variables over 1000 runs of collective infestations for both scenarios. One of the features that stands out is that in scenario 2 the presence of the individual with the much larger threshold makes the collective infestation much more severe, in terms both of duration and number of lice. Thus, this individual acts as what in epidemiology is known as a `superspreader'~\cite{lloyd}) and so, for want of a better term, in the following we use this word to refer to this individual. One important aspect to notice is the difference between the average and the median of the infestation duration, that can be rather large. This happens because the distribution of infestation times is very skewed to the right, as can be seen from the histograms in Fig.~\ref{fighistos}.

\begin{figure} [!ht]
\centerline{\includegraphics[width=16cm,clip=true]{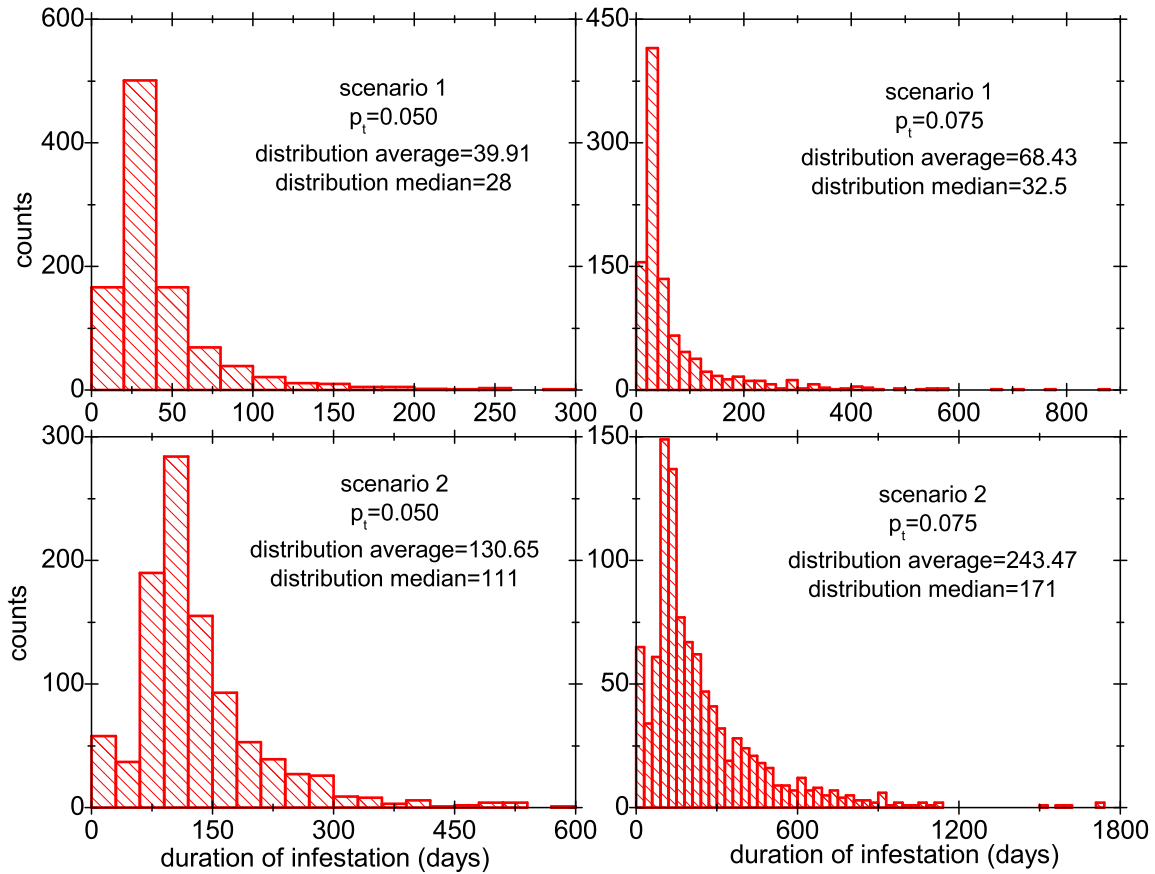}}
\caption{Histograms of the duration of the collective infestation for 1000 runs of scenario 1 (upper panels) and scenario 2 (lower panels) for two values of the probability of transfer: $p_t=0.05$ (left panels) and $p_t=0.075$ (right panels).} 
\label{fighistos}
\end{figure}

Interestingly the values shown in the last two rows of Table 4 are in general consistent with the values reported in a detailed study of lice infestations in several schools in Australia \cite{speare02}. Even though that study reports a somewhat higher number of lice per infested student than what our model predicts for $p_t=0.075$ or $p_t=0.1$, it must be bear in mind that in this section we have used several assumptions  that might not be correct for that population. For example, parents might be using different and less effective treatments from what we have assumed, and/or they might be starting to apply them with different thresholds. More importantly treatments are usually not applied in a systematic way, as has been assumed in this section. In this sense, it may be considered that ours are rather conservative scenarios.

\begin{figure} [!ht]
\centerline{\includegraphics[width=16cm,clip=true]{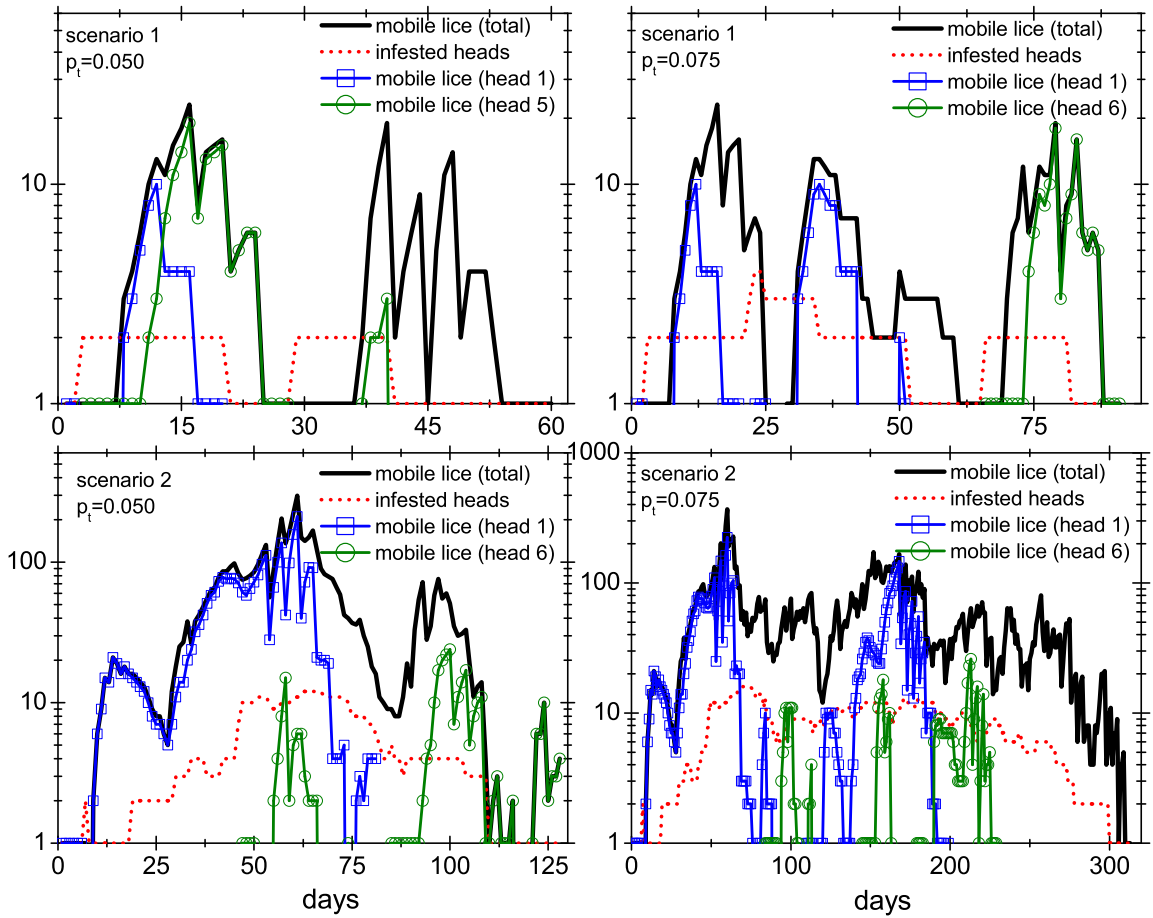}}
\caption{Time evolution of 4 different collective infestations in scenario 1 (upper panels) and scenario 2 (lower panels) for two values of the probability of transfer: $p_t=0.05$ (left panels) and $p_t=0.075$ (right panels). In all panels the total number of mobile lice as well as the number of infested heads is shown. For the sake of clarity, we only show the evolution of the infestation of two heads. Head 1 is the one where the first female louse appears, whereas the other is randomly chosen. Besides, for scenario 2 head 1 is also the superspreader head.} 
\label{figruns}
\end{figure}

There are some interesting features that stand out when single runs are considered (see Fig.~\ref{figruns}). One is that even in scenario 1, where the infestation is rather mild in terms of total number of mobile lice, re-infestations of the same head are not uncommon (see the evolution of mobile lice on head 1). For scenario 2, infestations are made more severe by the presence of the superspreader. Fig.~\ref{figruns} shows that scenario 2 infestations take a very long time to die out, even after the superspreader is no longer infested.

The histograms in Fig.~\ref{fighistos} show that in general infestations last a very long time, which is consistent with the widespread perception that head lice are very difficult to eradicate from a human group, even when individual actions are taken against the infestation. 

It is important to note that the simulations shown above only include the heads in a group, and it does not take into account the possible contagion from other members of the family of each child. 
To account for this, we allow the inclusion of external lice to an otherwise closed group of heads.
Our model shows that even a very weak `flux' of external lice, coming from family members or friends of the members of the group, can produce a significant increase in the duration of the collective infestation. As an example of this, Fig.~\ref{figcolados} shows the average (over 1000 simulations) duration of the collective infestation when one female adult louse is introduced in a random head every $n$ days in a population of $20$ heads (inside the colony the probability of transmission is  $p_t=0.075$, and the treatment used consists of applications of $80 \%$ efficacy every $4$ days). Note that the introduction of one new female louse every two weeks suffices to make the infestation last several months, or even years, even though all infested members use a systematic treatment. Interestingly, the increase in prevalence, defined as the proportion of infested heads, is much less significant reaching values that are still realistic.

\begin{figure} [!ht]
\centerline{\includegraphics[width=12cm,clip=true]{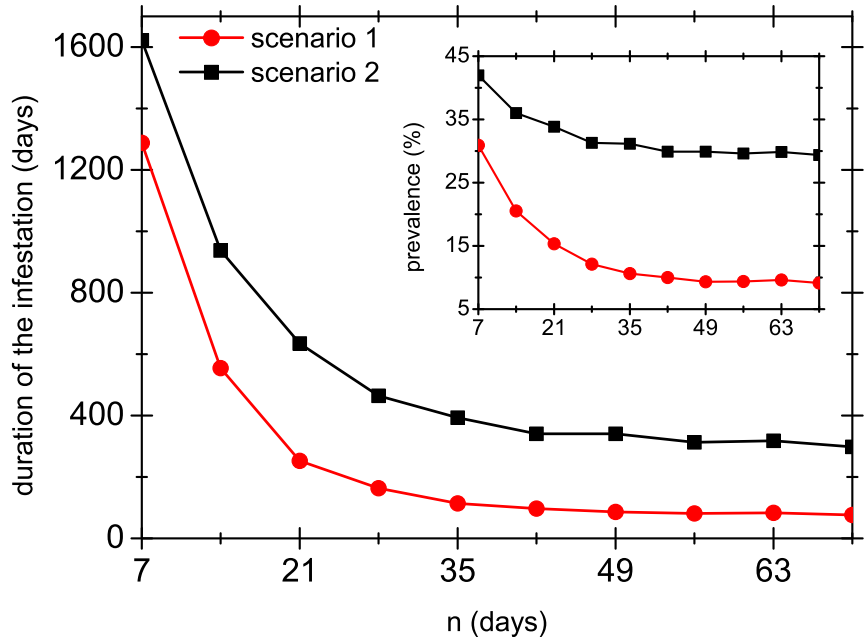}}
\caption{Average duration of infestation for a group of $20$ heads in which an external female adult louse is introduced every $n$ days. Transmission probability is $p_t=0.075$ and the treatment used consists of applications of $80 \%$ efficacy every $4$ days. Symbols represent average over $1000$ simulations (lines are guides to the eye). The inset shows the prevalence for the same situation, as a function of $n$.} 
\label{figcolados}
\end{figure}

Another important assumption is that each head is treated independently from other heads. But logic, as well as some experiences\cite{chouela,vermaak}, suggest that `synchronized' treatments might be effective in eradicating head lice from a human group, be it a school or a community. Synchronized treatment of a population means than when the number of mobile lice in one head becomes larger than a given threshold, the systematic treatment is begun in every head of the population. In other words, a parent that detects that his/her child has lice informs all the the other parents, who commit themselves to apply the same treatment on exactly the same days. In the unsynchronized case the parents act in isolation and only begin applying the treatment when lice are detected on his/her own child.

We have used our model to assess the performance of different treatments in a group of 3 heads (Fig.~\ref{figsyncnosync3}) and in a group of 20 heads (Fig.~\ref{figsyncnosync20}) with four different transmission probabilities $p_{t}$. We have chosen a systematic treatment applied every 4 days with an efficacy per application of $80 \%$. We compare the duration of the collective infestation of the synchronized case and the unsynchronized case in scenario 1.
The figures show that in this last case the duration of the treatment is multiplied by a constant (at least in the range of efficacies shown), and this constant gets larger as $p_t$ is increased. Intuitively, the picture is clear: some lice manage to avoid the application of the remedy by jumping from one head to another, and this gets worse with more mobile lice and larger groups of heads. On the other hand, when the treatment is applied at the same moment in the whole group of heads there is almost no dependence on the rate of transmission, because jumping from head to head does not help lice to avoid the treatment. But the model allows us to quantify these effects. For instance, the figures show that for $p_{t}=0.075$ the lack of synchronization increases the duration of the infestation of 3 heads by almost 50 \%, whereas for a group of 20 heads lack of synchronization almost doubles the duration of the infestation. It must be stressed that we are comparing with scenario 1, which lacks superspreaders. If these are included, the effect of synchronization is much more dramatic. To have an idea of the `perceived' duration of the infestation one must subtract the time it takes the population of lice to achieve the detection threshold, which is approximately 3 weeks (see inset of Fig 3).

\begin{figure} [!ht]
\centerline{\includegraphics[width=10cm,clip=true]{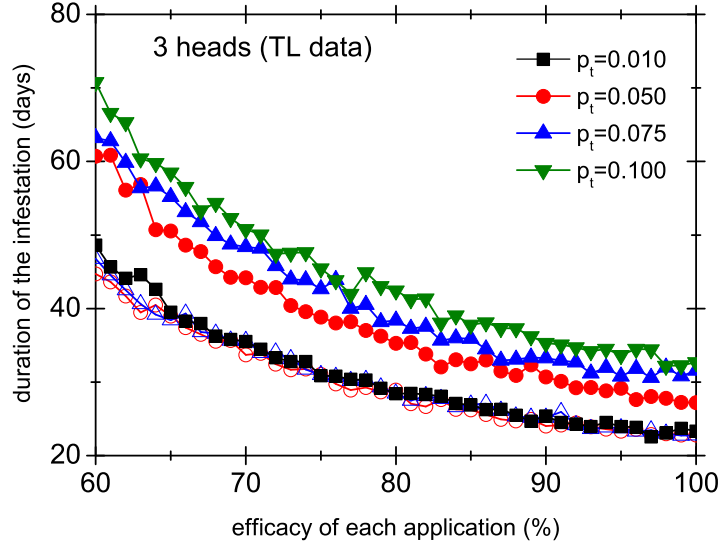}}
\caption{Comparison between the duration of the infestation in a group of 3 heads, for a systematic treatment applied every 4 days ($\Delta t_a=4$) applied in a synchronized (open symbols) and unsynchronized way (full symbols), , for different values of the transmission probability $p_t$ as a function of the efficacy of each application. 
For the unsynchronized case we have used scenario 1. } 
\label{figsyncnosync3}
\end{figure}

\begin{figure} [!ht]
\centerline{\includegraphics[width=10cm,clip=true]{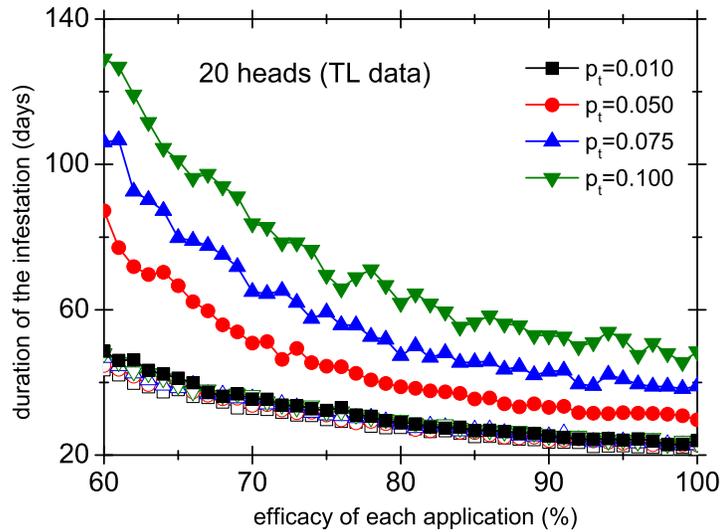}}
\caption{Comparison between the duration of the infestation in a group of 20 heads, for a systematic treatment applied every 4 days applied in a synchronized (open symbols) and unsynchronized way (full symbols), for different values of the transmission probability $p_t$ as a function of the efficacy of each application. 
For the unsynchronized case we have used scenario 1.} 
\label{figsyncnosync20}
\end{figure}

It is instructive to compare the effect of applying systematic versus non systematic strategies in groups of colonies. As in the previous section, the non systematic treatments cease to be applied when the number of mobile lice is below a threshold $\lambda_{end}$. Note that, by definition, these treatments are not synchronized because the precise days when the mobile lice number threshold is reached is an stochastic event. In Figs.~\ref{fig4treat3} and \ref{fig4treat20} we compare systematic treatments (synchronized and unsynchronized) with treatments that have a threshold of 1 and 2 mobile lice. The figures show that the difference between systematic and non-systematic treatments can be very large, even in the case that the former are not synchronized. For instance, with an application efficacy of 80 \% non-systematic treatments can last more than three times than the corresponding systematic treatment.

\begin{figure} [!ht]
\centerline{\includegraphics[width=12cm,clip=true]{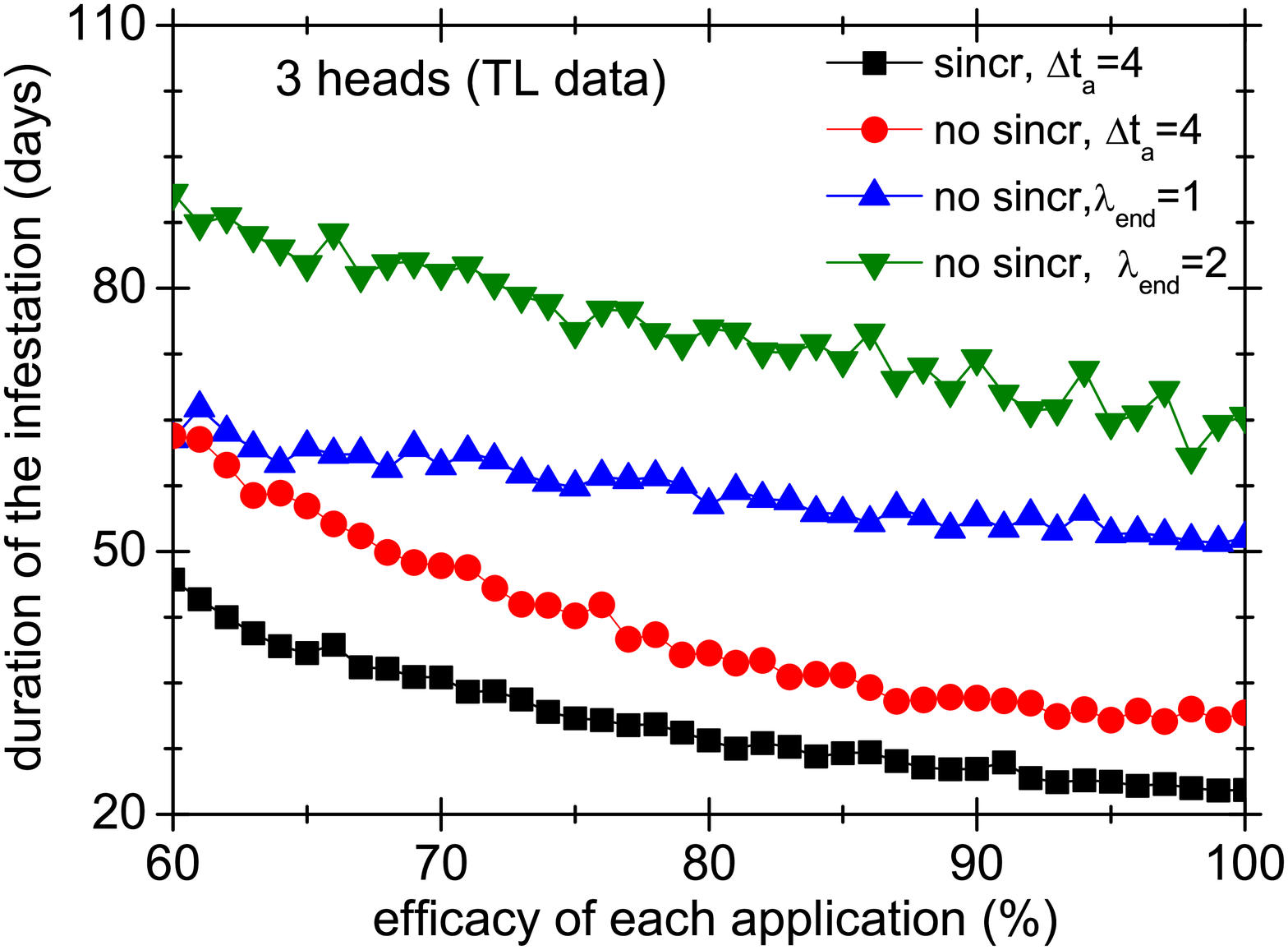}}
\caption{Comparison between the average duration over 1000 realizations of 4 different treatments in group of 3 heads, as a function of the efficacy of each application. Squares and circles correspond to systematic treatments, whereas triangles represent non-systematic ones.} 
\label{fig4treat3}
\end{figure}

\begin{figure} [!ht]
\centerline{\includegraphics[width=12cm,clip=true]{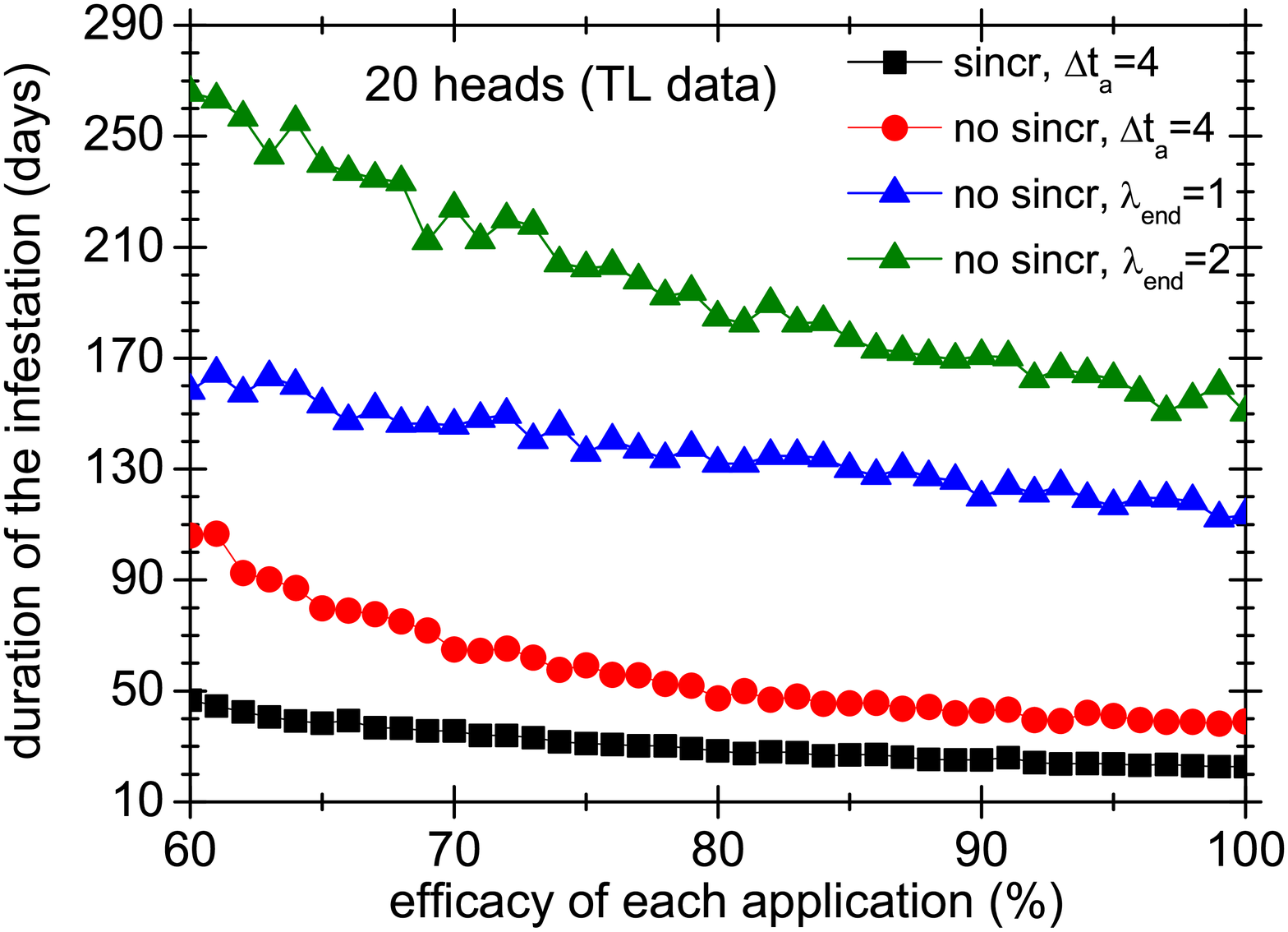}}
\caption{Comparison between the average duration over 1000 realizations of 4 different treatments in group of 20 heads, as a function of the efficacy of each application. Squares and circles correspond to systematic treatments, whereas triangles represent non-systematic ones.} 
\label{fig4treat20}
\end{figure}

\section*{Discussion}

We have presented in the previous sections a model of the evolution of populations of head lice based on detailed data about their biology. It has allowed us to address questions related to the `natural' growth of a population, as well as to assess the effectiveness of treatments that consist on several applications, as a function of the efficacy of each application. To explain the fact that colonies of head lice typically seem to be composed of a few insects, in spite of their naturally exponential growth, the best candidate seems to be self grooming of the host. But, whatever the mechanism, it is not clear what exactly would be its action on individual lice. Using our model we find that the reproductive success of female louse should be drastically reduced to achieve a significant slow down of population growth. On the other hand, a modest increase of the mortality of the various life stages (as it is the case with grooming) can lead to a very slow growth or even to the extinction of the population.

As is well known, a huge number of treatments have been proposed to deal with the problem of pediculosis. But, in general, the prediction of its outcome is either based on a few test cases, or in reasonings that take into account a few loose estimates of some aspects of the biology of the louse, and even in these cases, only very simple scenarios are analyzed. For example, assuming the treatment to be $100 \%$ effective against head lice, in \cite{lebwohl} the necessary application frequency for eradication is determined. But even though in vitro studies seem to suggest that the efficacy of some products is close to that value, it must be acknowledged that applying the product in a real head, in a real situation, is very likely to significantly reduce that efficacy. Thus, it is important to have a model where the result of a treatment can be studied as a function of the efficacy of each application. For the same example we find that even though 2 to 3 applications should be enough for a perfectly pediculicidal treatment, if the efficacy of each application is reduced to $80 \%$ the number of necessary applications increases to between 5 and 6 applications.

The model also allows the analysis of situations where a same treatment is applied in several different ways. As an example, we have analyzed what happens when treatments are applied systematically as compared with the results of non systematic applications. Needless to say, when systematicity is dropped, the exponential growth of possible treatments forces us to restrict to documented cases, or to common sense. For the treatments analyzed we find that systematicity can be very important: if the treatment is stopped when there remains only one adult lice (presumably because it cannot be detected), the whole duration of the treatment almost doubles the duration for the systematic variant. 
The problem with systematic treatments is that one would not know when to stop, because it is almost impossible to detect every lice and egg in a human head, and therefore it is impossible to know when the infestation has been completely eradicated. But it is exactly in these cases that models like ours can be useful, because they give a prediction of how many applications are necessary. In principle the model gives only average values, but the width of the distribution of number of applications is not large. Thus, using one more application than the average predicted by the model should ensure the eradication of lice in most cases.

To model the `contagion' of pediculosis among members of the same group, we have used the simplest variant: at all times every louse can jump to any other head, with a probability per day $p_t$. Even though this is a very simplified picture of reality (it lacks the possible formation of playing subgroups among children, does not take into account the likely differences in mobility for the various life stages of the louse, etc.) the model seems to capture the essence of the process, because the results obtained are compatible with what is found in the literature. We find that even when every individual applies a treatment to try to eradicate head lice, the duration of the collective infestation can be very large. The best strategy to completely eliminate a collective infestation in a reasonable time seems to be the synchronized application of the same treatment. 

Our model makes use of most of the data available about the biology of the head louse, what makes it dependent on a large number of parameters. In spite of this, we have shown that the predictions given are quite robust, because they are very similar for two different sets of data, one obtained by rearing head lice (TL) and the other obtained by rearing body lice (ES).

Evidently, even though our model is quite detailed, it could still be improved in many ways. For example, we have assumed that males are readily available and that with only one fertilization female lice are able to lay eggs until they die. This assumption could be dropped, but then one should add some form of interaction between male and female lice to the model, and have some idea of how many eggs can be laid after each fertilization. For some treatments, the efficacy could be assumed to depend on the life stage of the louse, and even on the number of lice present on the head which is being treated. Regarding contagion, the assumption of random mixing could be dropped, introducing a social network to model the interactions between children at school. Note however that many of these modifications would need to be based on real data that either are still very incomplete or do not even exist yet.

In a sense, our model is a compromise that tries to use as much detailed data as possible, while at the same time using reasonable assumptions for those processes that still are not well known. Yet, the level of detail used makes our model a useful tool to go beyond the usual educated guesses and predict the outcome of a large number of strategies in a number of different situations. In this sense, the cases analyzed in this paper are only examples of what can be done with the model. Its real strength lies in the possibility of adjusting it to analyze the practical strategies that are suggested to eradicate lice of real human groups in specific real contexts.


\section*{Acknowledgements}

We thank Guillermo Abramson for suggesting to us that building a model of populations of head lice could be interesting as well as useful.

\bibliography{article}

\newpage

\appendix

\renewcommand{\thefigure}{A\arabic{figure}}
\renewcommand{\thetable}{A\arabic{table}}

\section*{Appendix A}

In this appendix we provide details of the two sets of data used to build our models. We include additional tables constructed from the original data sets. We also give a brief description of the algorithm we use to perform the numerical simulations shown in the paper.

\subsection*{Description of data used for the models}

The data we have used for the models were taken from references $[8]$ and $[23]$ of the main text. The fraction of eggs that hatch a given number of days after oviposition are given in Table A1. These data do not include mortality of the whole egg stage, which was found to be $25 \%$ for Takano-Lee (TL) data and $12.3 \%$ for Evans-Smith (ES) ones. Table A2 gives the fraction of lice that undergo each moult a given number of days after hatching. Mortality is included in these sets of data and consequently the fractions for each moult do not add up to 1. The average mortality of the whole larvar stage is $25 \%$ (TL) and $13.7 \%$ (ES). As it is not possible to know how many lice or eggs die each day, we have used in our model a mortality rate of $3 \%$ per day (TL) and $1 \%$ per day (ES) for both the egg and larval stages. 
These values ​​were chosen so that the average mortality obtained from them is consistent with the average mortality data.

We have assumed that females lay a daily average of $5$ eggs, from their fourth day of adulthood until their deaths. For the first three days we have assumed that females lay an average of $0$, $2$ and $4$ eggs, respectively. This corresponds to the rise observed in egg production in both sets of data. Moreover, we have assumed that half of the eggs that hatch give rise to female lice, because no data set was reported to be sex-biased.

\begin{table}[ht!]
\centering
\caption{Fraction of eggs that hatch at a given day after oviposition, with respect to the total of eggs that hatch (i.e. eggs which do not hatch are not taken into account).}
\begin{tabular}{|c|c|c|c|}
\hline
\multicolumn{2}{|c|}{Takano-Lee}&\multicolumn{2}{c|}{Evans-Smith}\cr \hline
days&fraction&days&fraction \\
\hline \hline
$7$&$0.299$&$6$&$0.144$\cr \hline
$8$&$0.285$&$7$&$0.174$\cr \hline
$9$&$0.211$&$8$&$0.298$\cr \hline 
$10$&$0.127$&$9$&$0.305$\cr \hline
$11$&$0.078$&$10$&$0.079$\cr \hline
\end{tabular}
\label{table1}
\end{table}

\begin{table}[ht!]
\centering
\caption{Fraction of lice that moult at a given day after hatching. Fractions are expressed respect to the total of lice that have undergone the previous moult. Note that for some moults fractions do not add up to 1, reflecting the fact that some lice died before that moult.}
\begin{tabular}{|c|c|c|c|c|}
\hline
&\multicolumn{2}{|c|}{Takano-Lee}&\multicolumn{2}{c|}{Evans-Smith}\cr \hline
&days&fraction&days&fraction \\
\hline \hline
First Moult&$3$&$0.96$&$4$&$0.141$\cr \cline{2-5}
&$4$&$0.04$&$5$&$0.488$\cr \cline{2-5} 
&$$&$$&$6$&$0.308$\cr \cline{2-5}
&$$&$$&$7$&$0.026$\cr \hline
Second Moult&$5$&$0.673$&$8$&$0.413$\cr \cline{2-5}
&$6$&$0.154$&$9$&$0.512$\cr \cline{2-5} 
&$7$&$0.02$&$10$&$0.050$\cr \hline
Third Moult&$8$&$0.591$&$12$&$0.325$\cr \cline{2-5}
&$9$&$0.273$&$13$&$0.497$\cr \cline{2-5} 
&$$&$$&$14$&$0.141$\cr \hline
\end{tabular}
\label{table2}
\end{table}

For the ES data, the fraction of lice that are alive $x$ days after the last moult can be fitted by a Weibull function: $f(x)=\exp(-x/a)^2$, with $a=20.0$. Fig. \ref{figgrowth} shows that the fit is very good for female lice. The TL data are not so detailed and only give average and variance for the mortalities. We have assumed that the functional behaviour should be the same as for ES data and therefore we have simply looked for the value of $a$ which gives the average and variance of the Weibull function closest to the TL data: average mortality $20.2 \pm 1.4$. The value found is $a=22.8$ which gives an average of $20.2 \pm 2.3$. Note that the number of lice for which the TL average was obtained is relatively small: $19$ female lice.

A remark is in order regarding TL data. In their paper, the authors give the vital parameters of three {\it strains} of head lice, collected in three different places: California, Ecuador, and Florida. We have only used the data corresponding to the California strain because it has better statistics (more individuals for many life stages) and, more importantly, because the differences between strains are smaller than differences between the California strain data and ES data.

\begin{figure} [!ht]
\centerline{\includegraphics[width=12cm,height=9cm,clip=true]{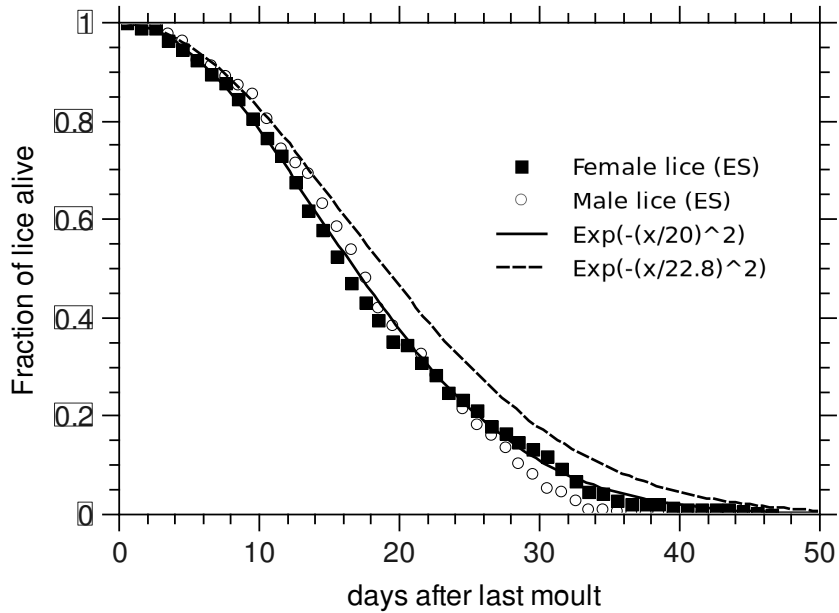}}
\caption{Fraction of lice alive as a function of the number of days elapsed since the last moult. Symbols correspond to ES data. The full curve is a fit to the curve of female lice mortality. The dotted curve gives the same average and variance of the mortalities in the TL data.} 
\label{figgrowth}
\end{figure}

\subsection*{Simulations}

As there are many relevant variables that cannot be calculated with the theoretical approach, we have resorted to stochastic simulations. This also allow us to see what happens in single infestations. In the following we give a brief description of the algorithm used.

All simulations start with an adult female louse that has moulted 10 days before. The number of days that she will live is a random number drawn from the corresponding mortality distribution (mentioned in the previous subsection). In a `day' of computation the following steps are followed:\\
1) {\it grooming}: each adult and nymph is killed with a given probability ($.05$ in all cases).\\
2) {\it nymph and egg mortality}: each nymph and egg is killed with a probability given probability ($.01$ for ES data and $.03$ for TL ones).\\
3) {\it births}: for each alive female adult lice a random integer number is drawn to know how many eggs she will lay. This depends on her age:\\
	-adults less than $3$ days old lay no eggs\\
	-3-day olds lay $1$, $2$, or $3$ eggs with probabilities $0.25$, $0.5$, $0.25$, respectively.\\
	-4-day olds lay $3$, $4$, or $5$ eggs with probabilities $0.25$, $0.5$, $0.25$, respectively.\\
	-adults older than $4$ days lay $4$, $5$, or $6$ eggs with probabilities $0.25$, $0.5$, $0.25$ respectively.\\
Each egg is assigned $5$ random integer numbers corresponding to the day each of the life stages will begin: hatching, first moult, second moult, third moult, and death. Every number is drawn from the corresponding distribution (see previous subsection). The internal clock of each new egg is set to $0$.\\
4) {\it death}: all adults whose internal clock reaches the established date of death are killed. \\
5) {\it contagion}: each alive female adult louse jumps to a randomly chosen head with a probability $p_t$. \\
6) {\it treatment}: each adult and nymph is killed with a probability given by the pediculicidity of the application, and eggs are killed with the corresponding ovicidal probability.\\
7) {\it counting}: the number of alive lice in each life stage is counted.\\
8) {\it updating}: the internal clock of each alive louse is incremented in one day.

In the oviposition step we have chosen a set of probabilities for the numbers of eggs layed, to add some stochasticity to the process. But we have checked that, given that the average is conserved, results depend only very weakly on these probabilities.

\section*{Appendix B}

\renewcommand{\thefigure}{B\arabic{figure}}
\setcounter{figure}{0}
\renewcommand{\thetable}{B\arabic{table}}
\setcounter{table}{2}

In the main text, most tables and figures are made ​​only with the TL data, to avoid a multiplication of figures and tables. In this appendix we provide for each of them in the main text its analogue built from ES data. To make the comparison easier, figures and tables have the same number as their analogue in the main text, but preceded by a letter B.

\begin{figure} [!ht]
\centerline{\includegraphics[width=12cm,clip=true]{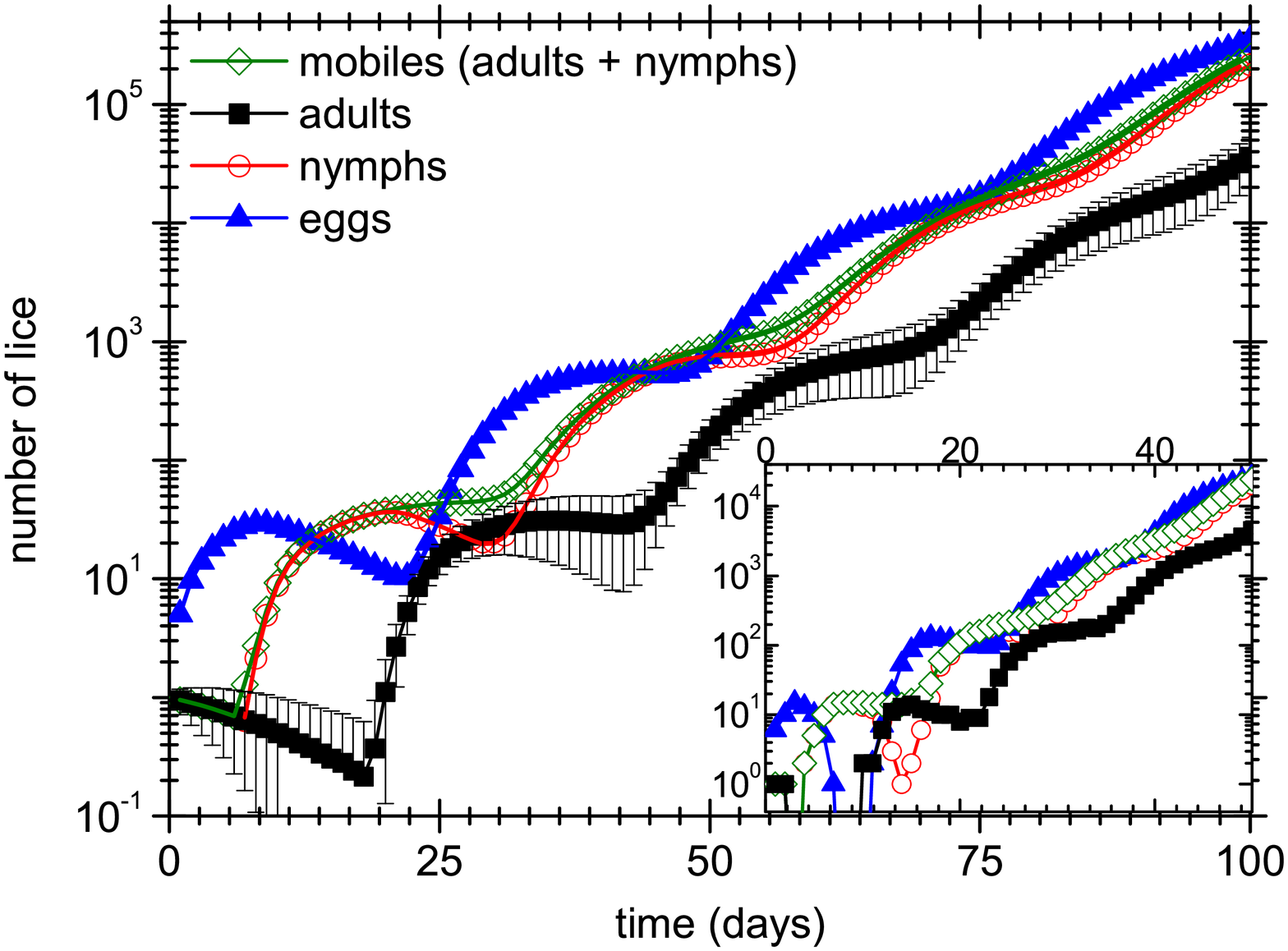}}
\caption{Average number of lice of a colony that is started at day $0$ by a female that had her last moult 10 days before. Symbols represent averages taken over $1000$ populations whereas full lines represent the theoretical predictions. The inset shows the first days of one of these populations. Here, as in the rest of the figures, the error bars represent the standard error of the mean. Data whose error bars are not shown have standard errors lower than symbol size.} 
\label{fignatevol}
\end{figure}

\setcounter{figure}{2}

\begin{figure} [!ht]
\centerline{\includegraphics[width=15cm,height=10cm,clip=true]{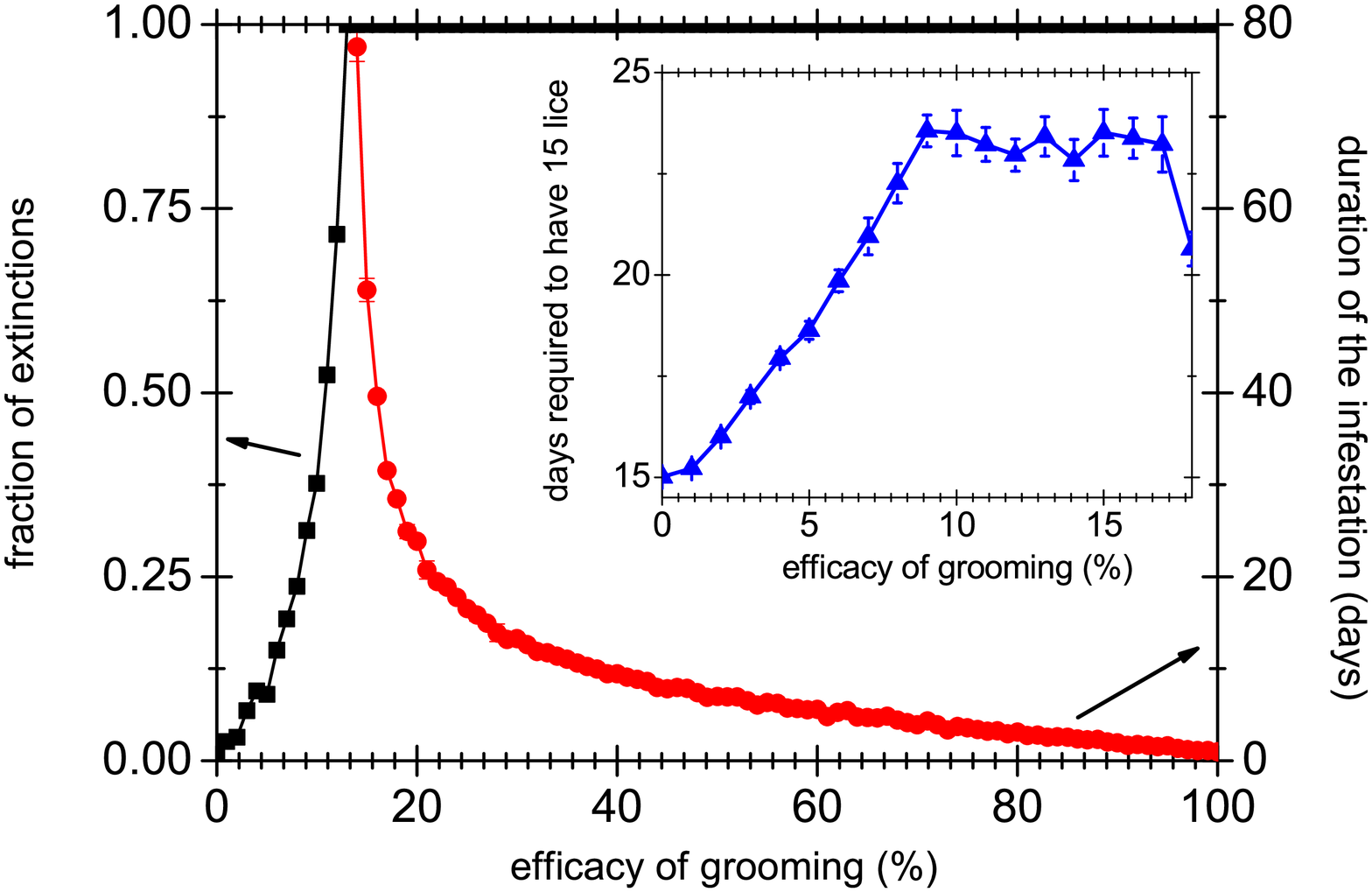}}
\caption{Fraction of extinctions (squares), and average duration of the infestation for the extinct populations (circles), as a function of the efficiency of grooming. Averages were taken over 1000 realizations, with a limit time of 500 simulation days (i.e. we only count extinctions happening within the first 500 days). The critical grooming efficacy is defined as the value at which the fraction of extinctions reach the unity. In this plot, this happens for efficacies closer to  $15 \%$. Inset: Number of days required to reach a population of $15$ lice as a function of the efficacy of grooming.} 
\label{figgroomin}
\end{figure}

\begin{figure} [!ht]
\centerline{\includegraphics[width=10cm,clip=true]{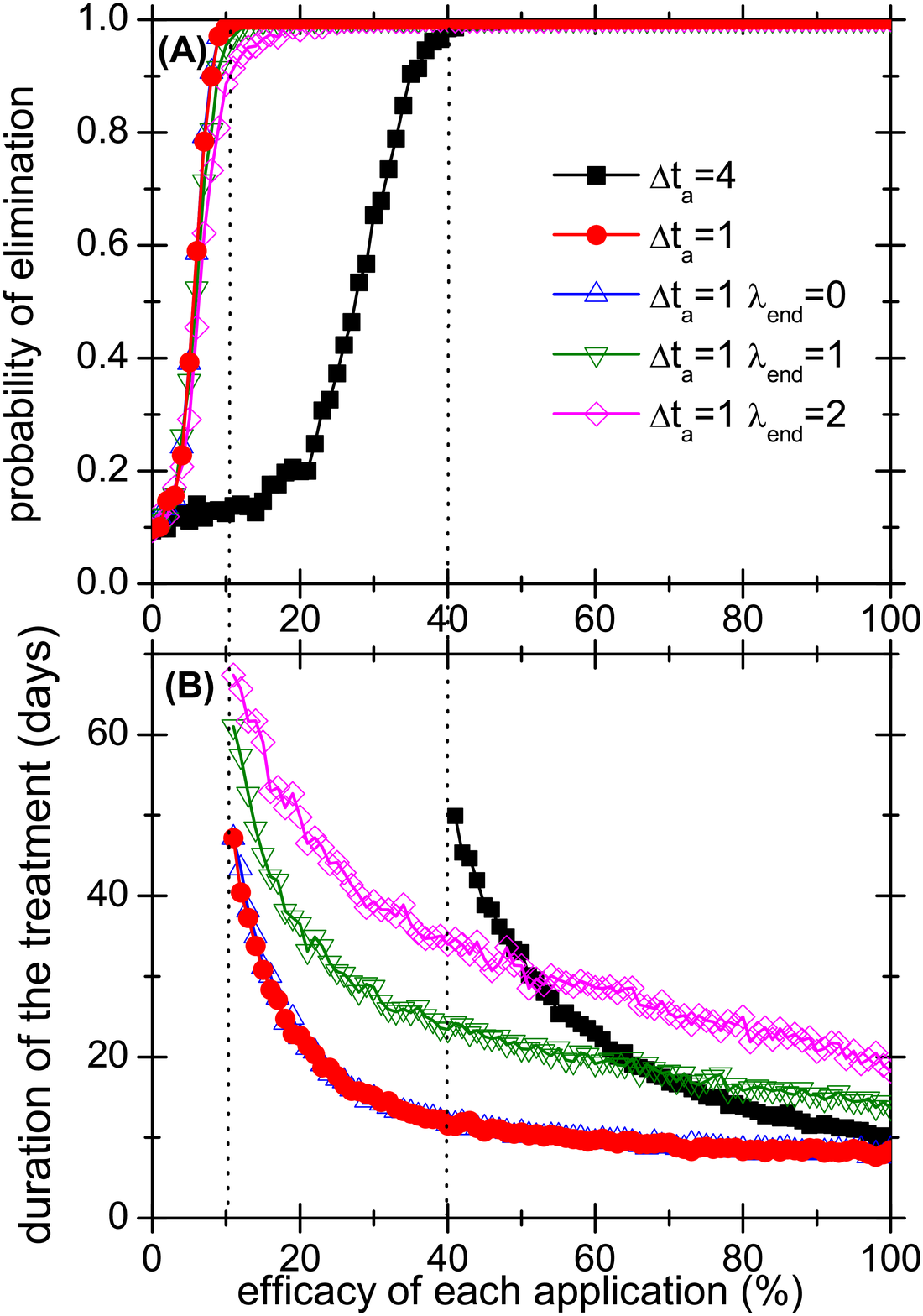}}
\caption{Results of applying different treatments to cure a head lice infestation. The upper panel shows the probability that the infestation is cured (i.e. that all head lice and eggs are eliminated) and the lower panel shows the duration of the treatment, when it is successful. Both variables are plotted as functions of the fraction of lice that are eliminated by each application of the treatment for a fixed ovicidity of $10\%$. The limit time in our simulations to allow for the extinction of the colony was $500$ days of simulation time. Dotted vertical lines indicate the critical efficacy of the treatments.} 
\label{figprobdur}
\end{figure}

\begin{figure} [!ht]
\centerline{\includegraphics[width=10cm,clip=true]{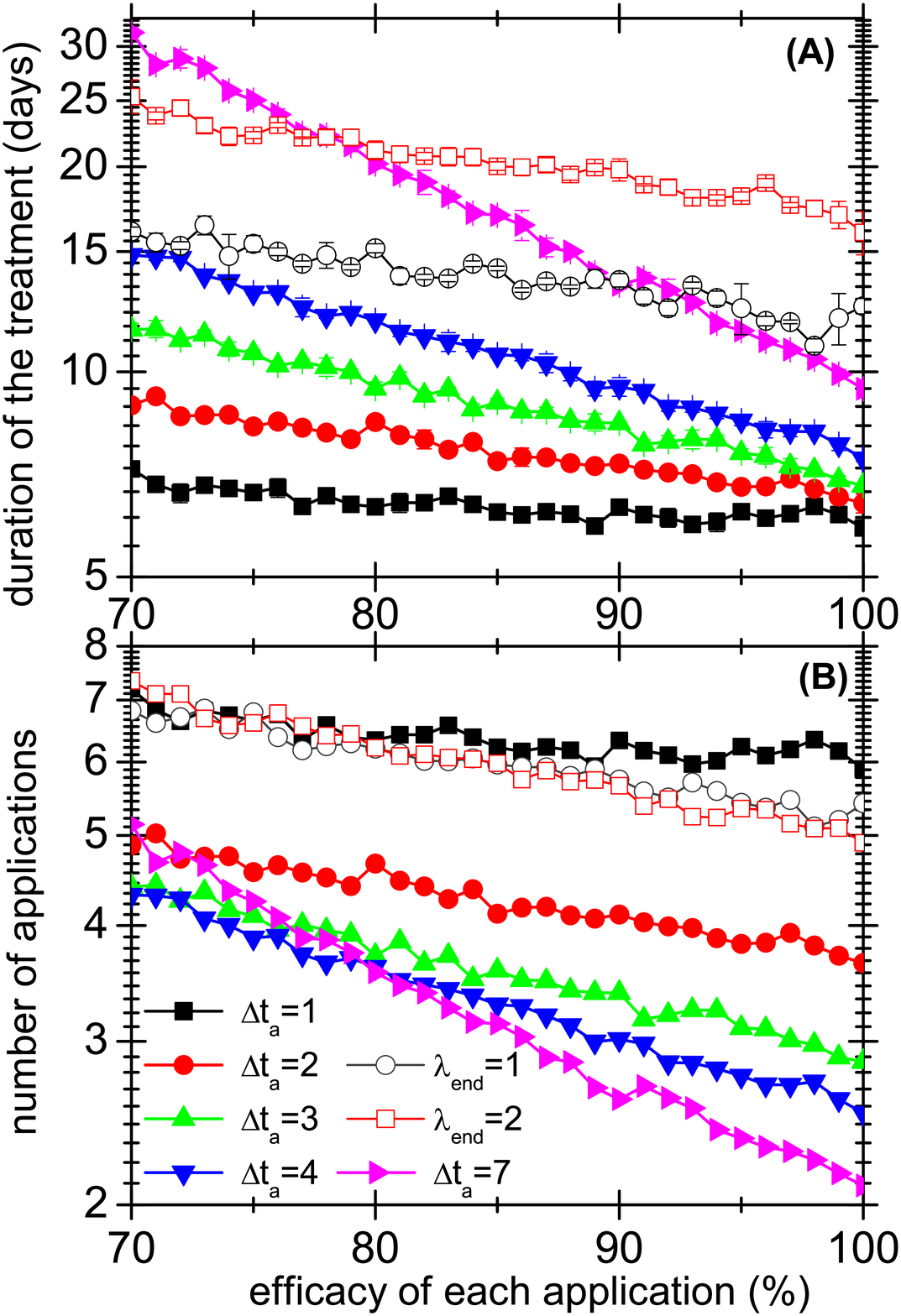}}
\caption{Comparison between the average durations (panel A) and between the number of applications (panel B) for several treatments, when they are successful, as a function of the fraction of lice eliminated by each application. Full symbols indicate systematic treatments, and $\Delta t_{a}$ gives the number of days between applications. Empty symbols correspond to a daily application ($\Delta t_a=1$) of non systematic treatments which are stopped when the mobile lice remaining in the population are less of equal than $\lambda_{end}$.} 
\label{figcompatreat1}
\end{figure}

\begin{figure} [!ht]
\centerline{\includegraphics[width=10cm,clip=true]{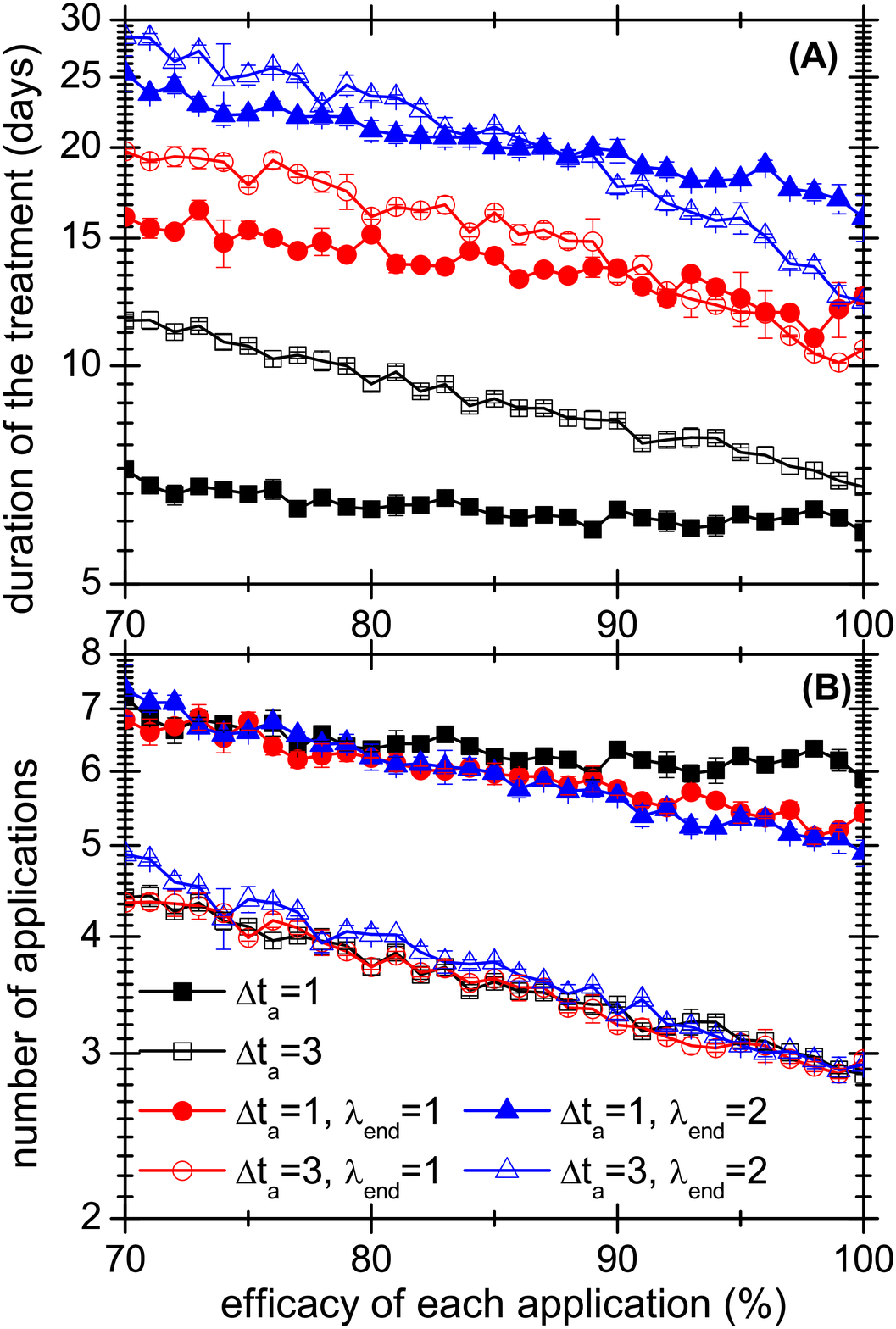}}
\caption{Comparison between the average durations (panel A) and between the number of applications (panel B) for several treatments, when they are successful, as a function of the fraction of lice eliminated by each application. Squared symbols are systematic treatments, whereas the rest are non systematic ones. Full symbols correspond to daily applications that are stopped when less than $\lambda_{end}$ mobile lice remain in the population. Empty symbols correspond to an application every 3 days.} 
\label{figcompatreat2}
\end{figure}

\begin{table}[ht!]
\centering
\caption{Result of the average of several variables over 1000 runs of collective infestations for scenario 1 (group of 20 heads with thresholds for treatment beginning randomly chosen between 10 and 20 mobile lice). In the first column the probability of transfer is indicated. Second column gives the average total daily number of mobile lice involved. The third is the prevalence, i.e., the proportion of infested heads. Fourth column indicates the average daily number of lice transferred from one head to another. The fifth and sixth columns give the average and median duration of collective infestations, respectively.}
\begin{tabular*}{\textwidth}{@{\extracolsep{\fill}}|c|c|c|c|c|c|}
\hline
$p_t$&total&prevalence&total&average duration&median duration\cr
&mobile&&transferred&of infestation&of infestation\cr
\hline \hline
$0.010$&$3.67$&$4.57 \%$&$0.095$&$27.11$&$24$ \cr \hline
$0.050$&$4.48$&$5.78 \%$&$0.82$&$34.22$&$24\renewcommand{\theequation}{A-\arabic{equation}}
$ \cr \hline
$0.075$&$5.43$&$6.94 \%$&$2.53$&$46.41$&$27$ \cr \hline
$0.100$&$6.97$&$8.63 \%$&$6.60$&$63.94$&$33$ \cr \hline
\end{tabular*}
\label{table3}
\end{table}

\begin{table}[ht!]
\centering
\caption{Result of the average of several variables over 1000 runs of collective infestations for scenario 2 (almost identical to scenario 1, but with one individual having a treatment threshold of 100 mobile lice). In the first column the probability of transfer is indicated. Second column gives the average total daily number of mobile lice involved. The third is the prevalence, i.e., the proportion of infested heads. Fourth column indicates the average daily number of lice transferred from one head to another. The fifth and sixth columns give the average and median duration of collective infestations, respectively.}
\begin{tabular*}{\textwidth}{@{\extracolsep{\fill}}|c|c|c|c|c|c|}
\hline
$p_t$&total&prevalence&total&average duration&median duration\cr
&mobile&&transferred&of infestation&of infestation\cr
\hline \hline
$0.010$&$22.77$&$6.55 \%$&$1.40$&$72.59$&$73$ \cr \hline
$0.050$&$25.52$&$14.35 \%$&$12.53$&$112.13$&$107$ \cr \hline
$0.075$&$29.36$&$19.65 \%$&$30.89$&$161.02$&$139$ \cr \hline
$0.100$&$35.54$&$25.8 \%$&$71.80$&$231.59$&$182$ \cr \hline
\end{tabular*}
\label{table4}
\end{table}

\begin{figure} [!ht]
\centerline{\includegraphics[width=16cm,clip=true]{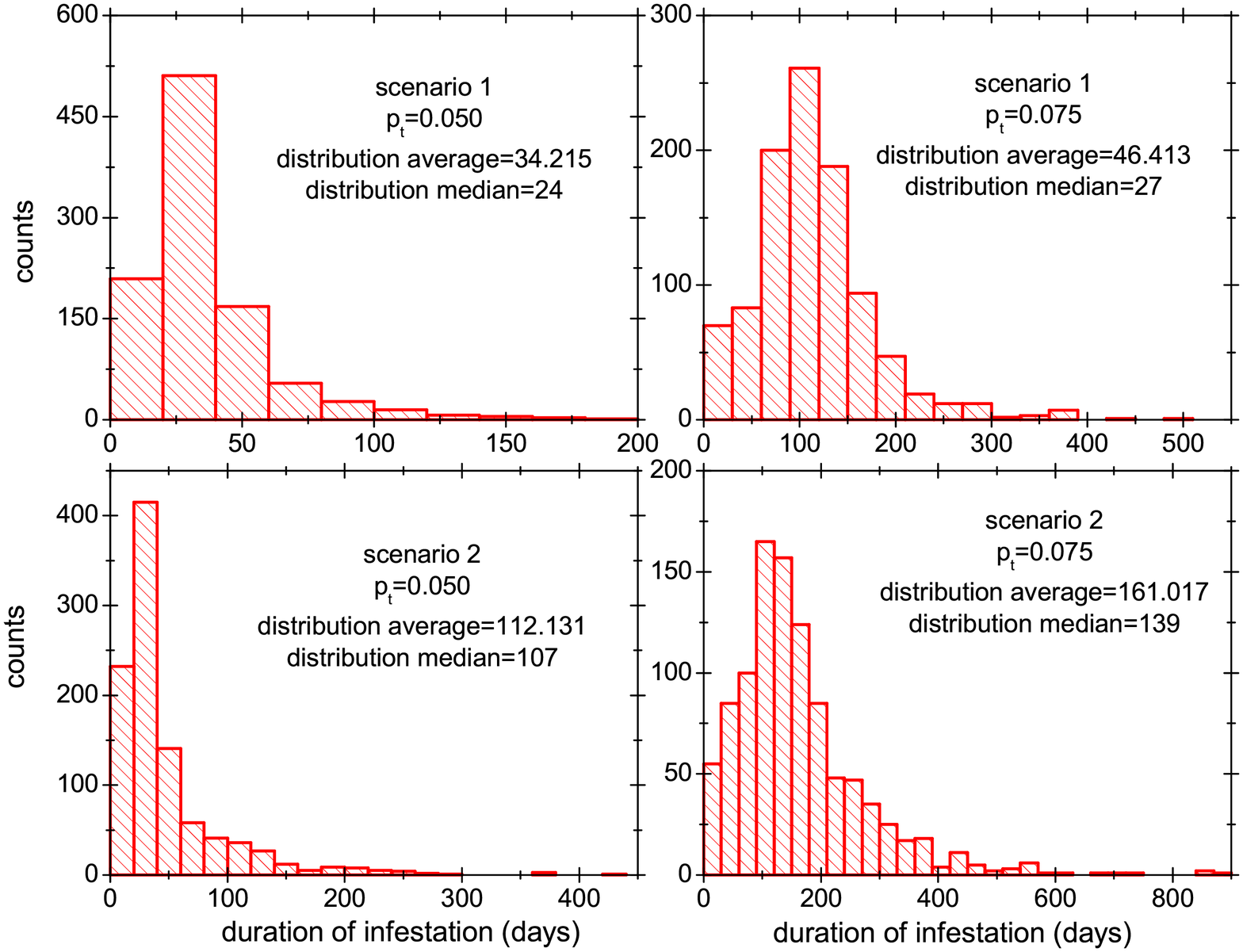}}
\caption{Histograms of the duration of the collective infestation for 1000 runs of scenario 1 (upper panels) and scenario 2 (lower panels) for two values of the probability of transfer: $p_t=0.05$ (left panels) and $p_t=0.075$ (right panels).} 
\label{fighistos}
\end{figure}

\begin{figure} [!ht]
\centerline{\includegraphics[width=16cm,clip=true]{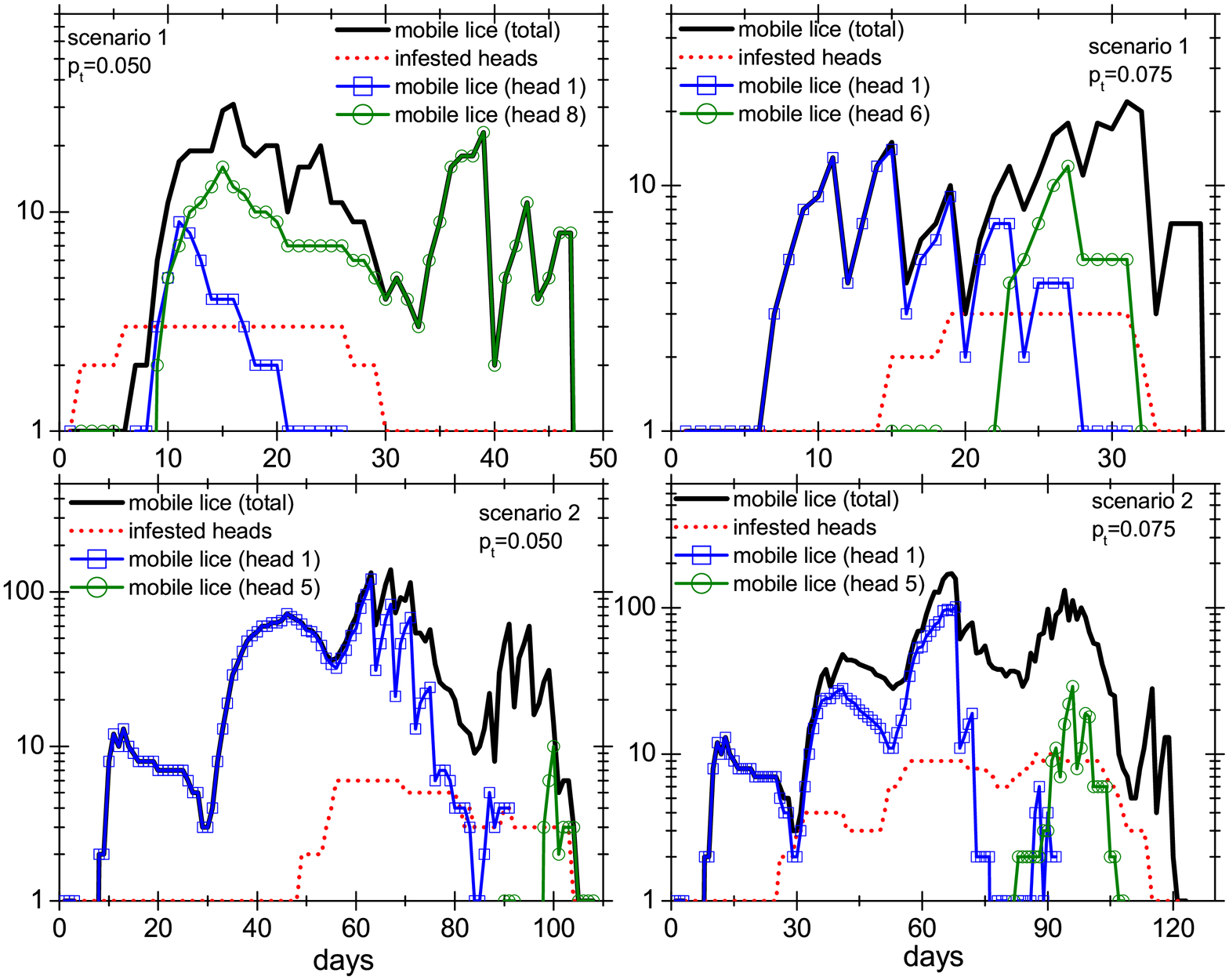}}
\caption{Time evolution of 4 different collective infestations in scenario 1 (upper panels) and scenario 2 (lower panels) for two values of the probability of transfer: $p_t=0.05$ (left panels) and $p_t=0.075$ (right panels). In all panels the total number of mobile lice as well as the number of infested heads is shown. For the sake of clarity, we only show the evolution of the infestation of two heads. Head 1 is the one where the first female louse appears, whereas the other is randomly chosen. Besides, for scenario 2 head 1 is also the superspreader head.} 
\label{figruns}
\end{figure}

\begin{figure} [!ht]
\centerline{\includegraphics[width=10cm,clip=true]{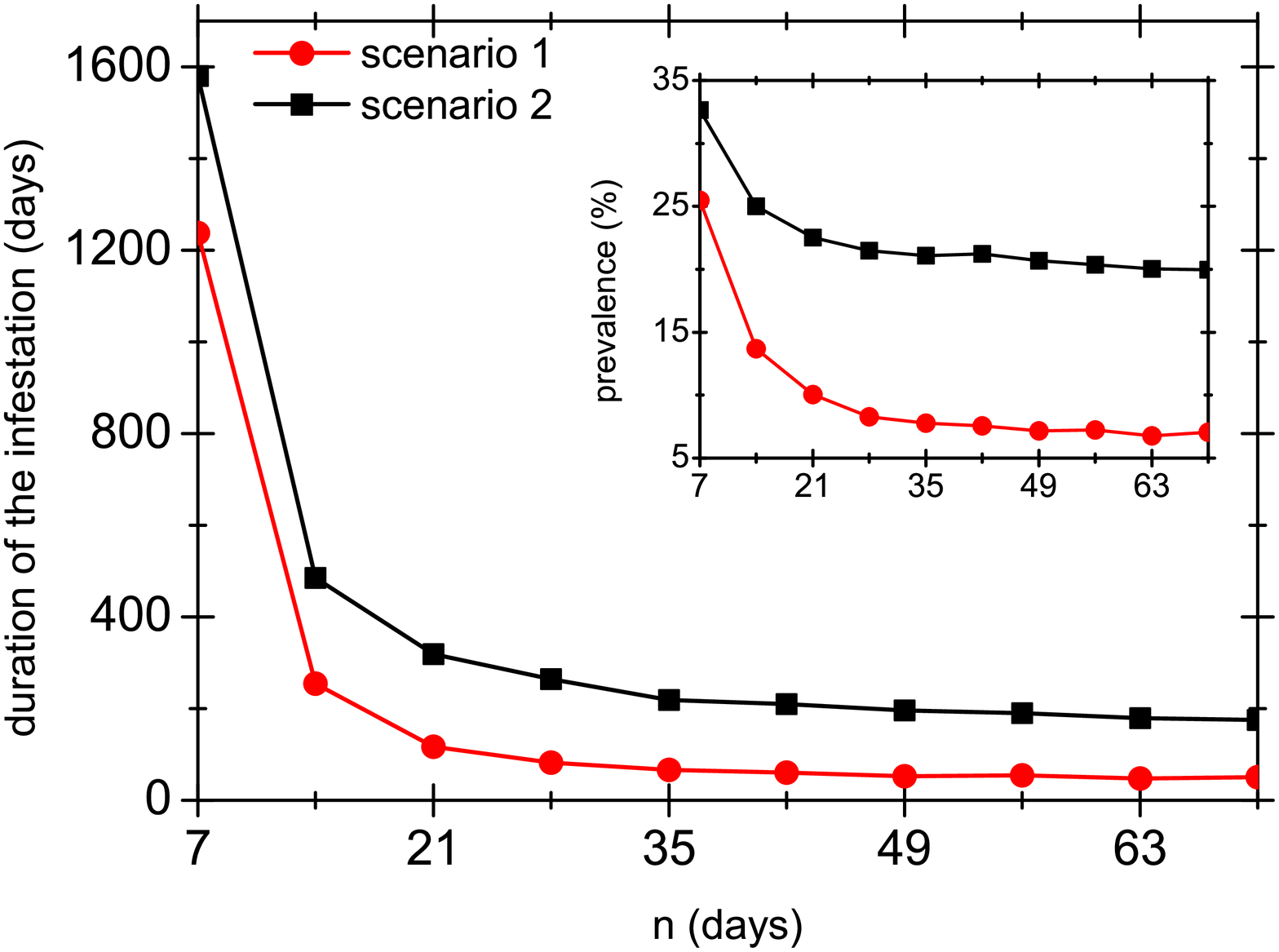}}
\caption{Average duration of infestation for a group of $20$ heads in which an external female adult louse is introduced every $n$ days. Transmission probability is $p_t=0.075$ and the treatment used consists of applications of $80 \%$ efficacy every $4$ days. Symbols represent average over $1000$ simulations (lines are guides to the eye). The inset shows the prevalence for the same situation, as a function of $n$.} 
\label{figcolados}
\end{figure}

\begin{figure} [!ht]
\centerline{\includegraphics[width=10cm,clip=true]{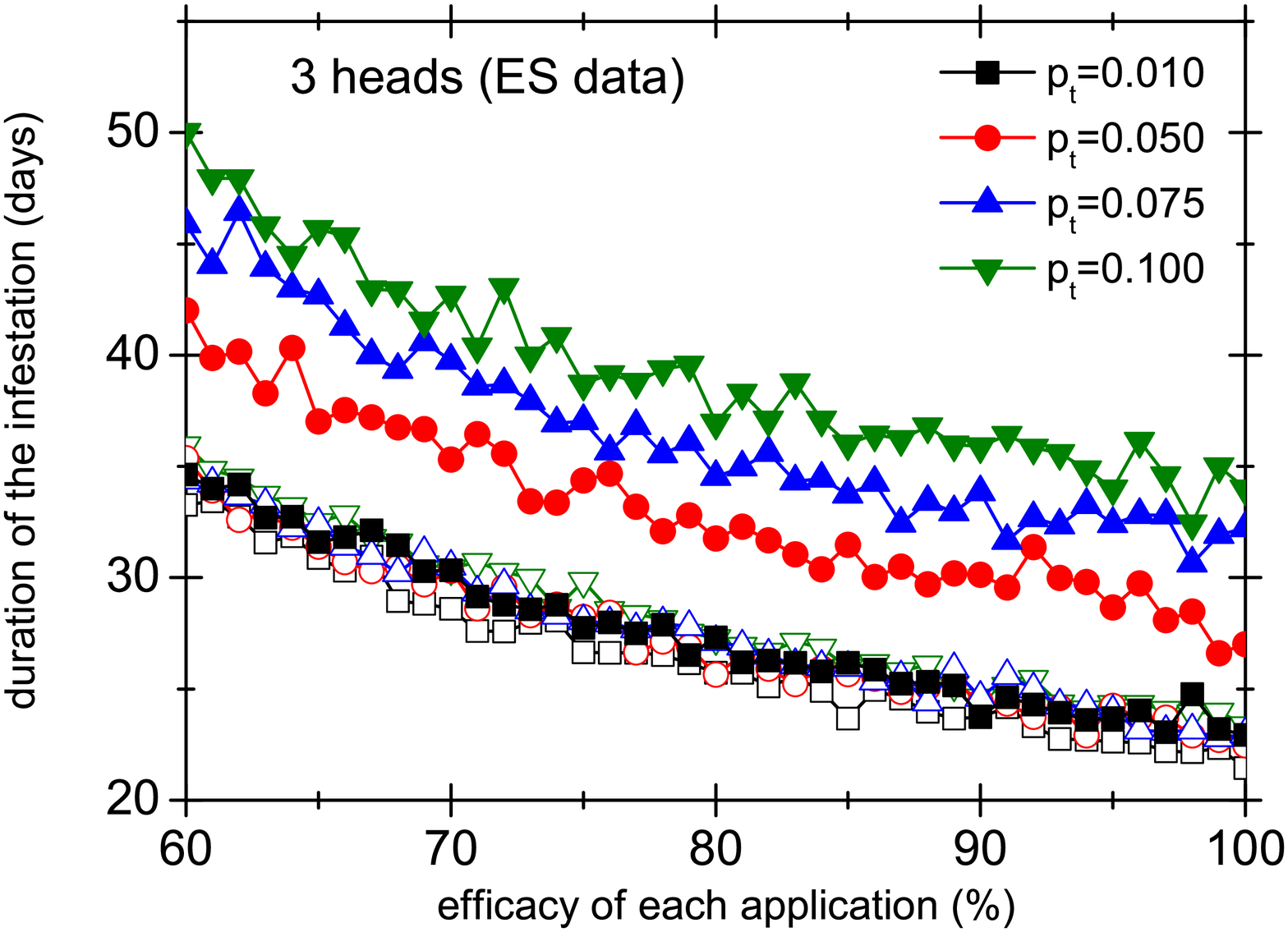}}
\caption{Comparison between the duration of the infestation in a group of 3 heads, for a systematic treatment applied every 4 days ($\Delta t_a=4$) applied in a synchronized (open symbols) and unsynchronized way (full symbols), , for different values of the transmission probability $p_t$ as a function of the efficacy of each application. 
For the unsynchronized case we have used scenario 1. } 
\label{figsyncnosync3}
\end{figure}

\begin{figure} [!ht]
\centerline{\includegraphics[width=10cm,clip=true]{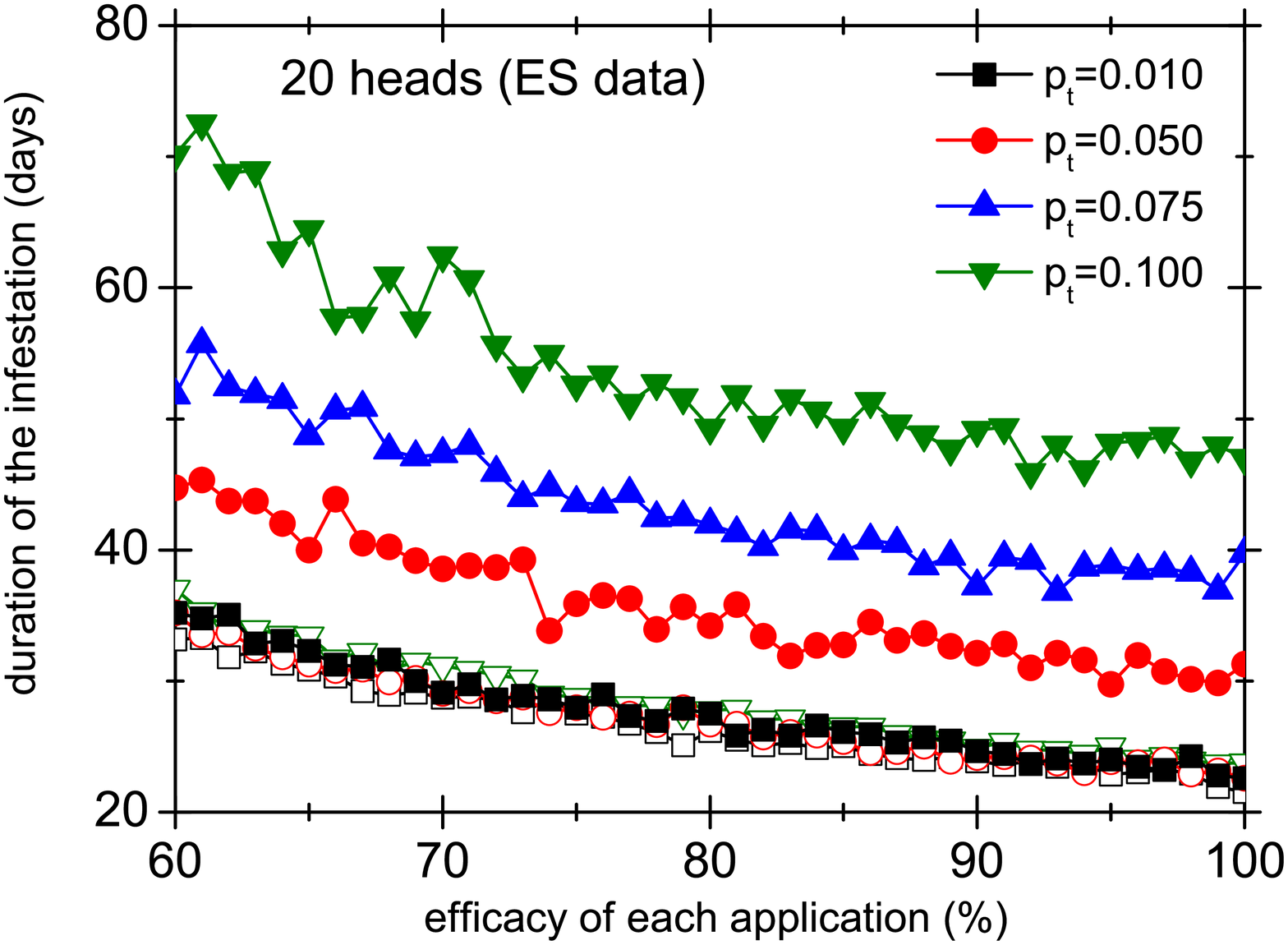}}
\caption{Comparison between the duration of the infestation in a group of 20 heads, for a systematic treatment applied every 4 days applied in a synchronized (open symbols) and unsynchronized way (full symbols), for different values of the transmission probability $p_t$ as a function of the efficacy of each application. 
For the unsynchronized case we have used scenario 1.} 
\label{figsyncnosync20}
\end{figure}

\begin{figure} [!ht]
\centerline{\includegraphics[width=10cm,clip=true]{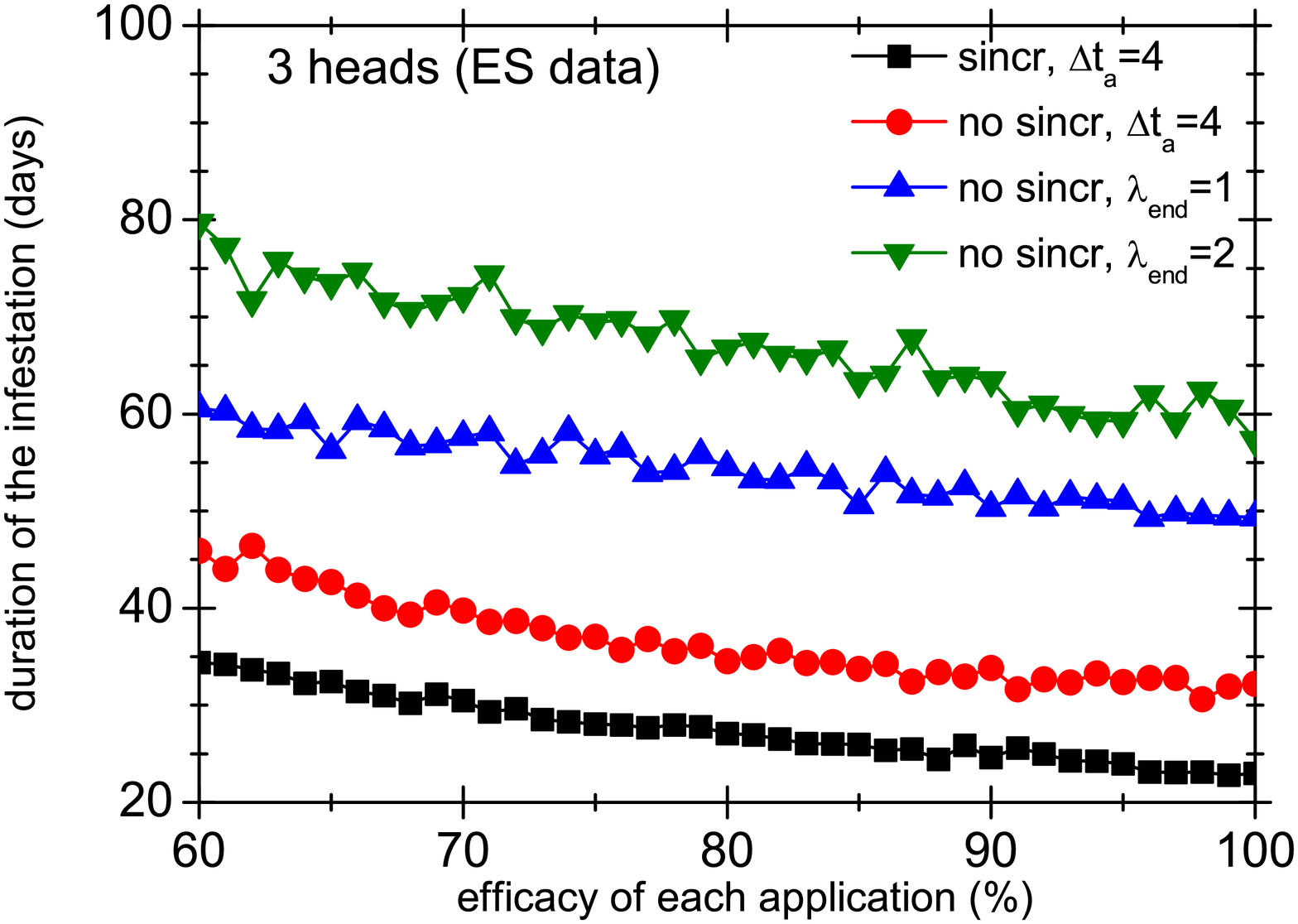}}
\caption{Comparison between the average duration over 1000 realizations of 4 different treatments in group of 3 heads, as a function of the efficacy of each application. Squares and circles correspond to systematic treatments, whereas triangles represent non-systematic ones.} 
\label{fig4treat3}
\end{figure}

\begin{figure} [!ht]
\centerline{\includegraphics[width=10cm,clip=true]{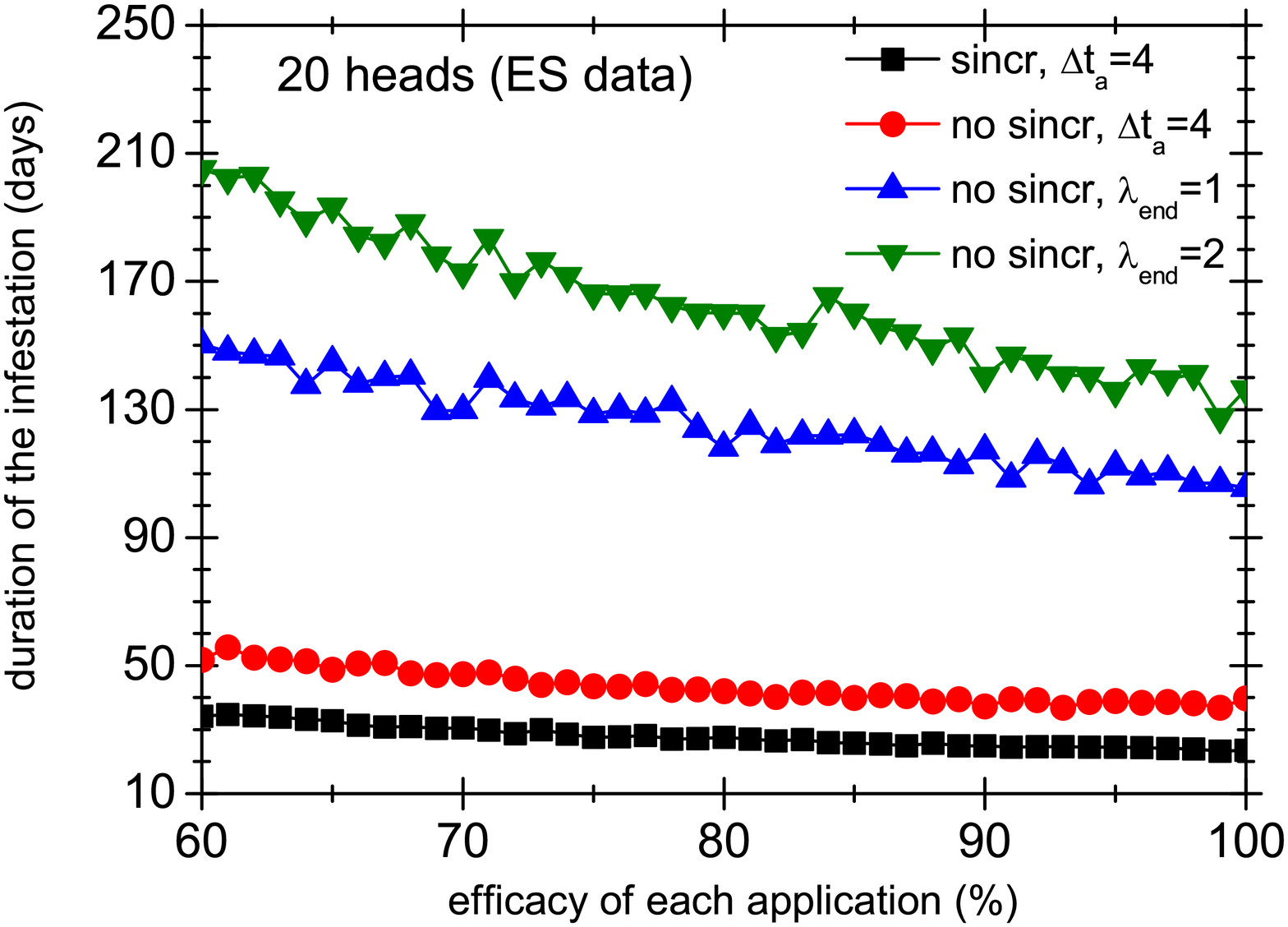}}
\caption{Comparison between the average duration over 1000 realizations of 4 different treatments in group of 20 heads, as a function of the efficacy of each application. Squares and circles correspond to systematic treatments, whereas triangles represent non-systematic ones.} 
\label{fig4treat20}
\end{figure}

\section*{Figure Legends}

\section*{Tables}

\end{document}